# entropy

*Article*

# Predicting Community Evolution in Social Networks

**Stanisław Saganowski** [1,†,]*, **Bogdan Gliwa** [2,†], **Piotr Bródka** [1,†], **Anna Zygmunt** [2,†],
**Przemysław Kazienko** [1,†] and **Jarosław Koźlak** [2,†]

1   Department of Computational Intelligence, Wrocław University of Technology,
    Wyb.Wyspiańskiego 27, 50-370 Wrocław, Poland; E-Mails: piotr.brodka@pwr.edu.pl (P.B.);
    kazienko@pwr.edu.pl (P.K.)
2   AGH University of Science and Technology, Al. Mickiewicza 30, 30-059 Kraków, Poland;
    E-Mails: bgliwa@agh.edu.pl (B.G.); azygmunt@agh.edu.pl (A.Z.); kozlak@agh.edu.pl (J.K.)

†   These authors contributed equally to this work.

*   Author to whom correspondence should be addressed; E-Mail: stanislaw.saganowski@pwr.edu.pl;
    Tel.: +48-71-320-36-09; Fax: +48-71-320-34-53.

Academic Editor: J. A. Tenreiro Machado

**Abstract:** Nowadays, sustained development of different social media can be observed
worldwide. One of the relevant research domains intensively explored recently is analysis
of social communities existing in social media as well as prediction of their future
evolution taking into account collected historical evolution chains. These evolution chains
proposed in the paper contain group states in the previous time frames and its historical
transitions that were identified using one out of two methods: Stable Group Changes
Identification (SGCI) and Group Evolution Discovery (GED). Based on the observed
evolution chains of various length, structural network features are extracted, validated and
selected as well as used to learn classification models. The experimental studies were
performed on three real datasets with different profile: DBLP, Facebook and Polish
blogosphere. The process of group prediction was analysed with respect to different
classifiers as well as various descriptive feature sets extracted from evolution chains of
different length. The results revealed that, in general, the longer evolution chains the better
predictive abilities of the classification models. However, chains of length 3 to 7 enabled
the GED-based method to almost reach its maximum possible prediction quality. For
SGCI, this value was at the level of 3–5 last periods.

**Keywords:** social network; social network analysis (SNA); social community; social group detection; group evolution; group evolution prediction; group dynamics; classifier; feature selection; GED; SGCI

## 1. Introduction

Social networks—regarded as social relationships connecting human entities—have become a more and more popular research domain, mainly due to a considerable interest in the Internet social media, such as Facebook, Twitter or the blogosphere. The most popular definition of a social network describes it as a finite set or multimodal sets of actors combined with the relation or relations defined on them [1].

Existing social media provide access to rich information about individuals—users of various services and different kinds of relationships between them. Obtaining such information and its detailed analysis may support the understanding of social processes taking place in the virtual world. The data should be properly acquired, stored and organised. This information can be represented by means of a dependency graph and analysed using complex network methodologies.

Social Network Analysis (SNA) utilizes methods coming from different domains such as graph theory, sociology, physics and computer science. Additionally, the real social networks are not homogenous—they consist of subareas where the density of internal connections is higher than external ones. Based on this observation, many algorithms for social groups' identification in networks have been proposed. The majority of them assume a static character of networks, thus, they provide static analysis of social groups. It appears, however, that such assumption significantly simplifies the real world, in which users of social media and connections between them continually change over time.

## 2. Problem Description

Social communities are evolving together with the changes of the entire network; they appear, disappear, merge, split as well as new members join or leave existing groups. Developing methods to track such group evolution makes it possible to understand the background and reasons ruling human behaviour, and to use them in facing with many practical problems that arise in marketing, politics or public security domains.

In the case of marketing, it may be related with the analysis of possible impact while introducing a new product or services, e.g., why some incentives reduce user interactions in web-based customer support services. In politics, it may embrace an observation of influences of given political programs or individual politicians on some social groups and the analysis of influence evolution in time. Particularly, it may be used to monitor collective reactions to the course of election campaign or to the introduction of changes in the law.

In public security affairs, the observation of the group evolution may facilitate identification of users or groups who propagate or support dangerous or criminal ideas and behaviour, e.g., terrorism.

Forecasting possible future behaviour of social groups based on the analysis of their history, would be another useful step after analysis of former group evolution. It may be considered as an advice suggesting taking adequate actions to counteract the predicted direction of evolution. Hence, the main problem addressed in this paper is the development of effective methods that could be used to predict the evolution of social communities existing in the temporal social network in the nearest future, *i.e.*, in the next period.

Nowadays, there exist some algorithms for the identification of group state evolution in time–transition type detection, but there are no studies on prediction of future changes of the group states (next events) as well as their indicators. It is a very complex problem since it is influenced by behaviour of individuals, which is difficult to predict. Moreover, the human reactions may be triggered by events or external information, which are not monitored and cannot be reflected in the social network.

At the initial stage of community evolution prediction, it is necessary to acquire and prepare the adequate data related to group evolution, *i.e.*, we need to identify groups in particular periods, to determine transitions between groups and event types in subsequent time slots.

To obtain a sufficient quality of prediction, it is essential to select a proper set of information (measures-features) characterising group states and historical events related to them. Additionally, it is necessary to select the most valuable features, to apply a suitable prediction method and to define the most appropriate history of group states–the number of time windows taken into consideration. It can be supposed that these choices should be interrelated to the general profile of group dynamics. It should be emphasized that no effective methods and algorithms to predict future changes of social communities have been proposed and deeply investigated so far.

## 3. Related Work

In recent years dozens of community extraction methods have been developed, also several methods to track changes of the group over the time have been presented. Lately, one of the most investigated aspect of social network analysis is prediction. The best investigated is link prediction problem [2–4]. It refers to predicting the existence of a link (relation) between two nodes (users) within a social network. Prediction is being made based on different network and group measures. For example Liben-Nowell *et al.* [2] focused on path and common neighbours between pair of nodes, while Lichtenwalter *et al.* [3] consider degrees and mutual information between them. Zheleva *et al.* [4] explored different combinations of descriptive, structural and group features (e.g., group membership) and proved that prediction accuracy is 15%−30% more accurate as compared to using descriptive node attributes and structural features.

After successful results in link prediction the researchers have immersed in the problem to link sign prediction [5–8]. Sign in this context means that predicted relation between users may be positive or negative. Again the prediction is being made based on network and group measures. Symeonidis *et al.* [8] looked at paths between the node pair and used the notion of similarity to predict the sign. Leskovec *et al.* [7] used degree and mutual information between pair of nodes for link prediction and profits from the theory of balance and status to predict the link sign. Kunegis *et al.* [6] evaluated different signed spectral similarity measures to predict the sign of the link in Slashdot.

Davis *et al.* [9] tackled the problem of multi-relational link prediction by extending the neighbourhood methods with weight and focusing on triads. Richter *et al.* [10] and Wai-Ho *et al.* [11]

faced the very important task of churn prediction. Wai-Ho *et al.* introduced a new data mining algorithm called DMEL (data mining by evolutionary learning), which estimates each prediction being made. Richter *et al.* presented a novel approach and tried to predict churn based on analysis of group behaviour. This approach touches another aspect, not well studied yet, where evolution of the whole group is being predicted, *i.e.*, which event will be next in group lifetime.

Some approaches regard forecasting of changes of groups in the future. Kairam *et al.* [12] investigated the possibility of prediction whether a community will grow and whether survive in the long term. They achieved a good accuracy—over 77% in predicting group growth over the following 2 months, 2 years and group death within a year. Patil *et al.* [13] addressed a similar problem—whether a group will disappear or will thrive in the future and they obtained accuracy over 90%. Goldberg *et al.* [14] handled a problem of prediction of lifespan of evolution for a group. They indicated that there is a correlation between the lifespan of a community and its structural properties from early stages of evolution. Interesting question is if the evolution of the whole social network structure (*i.e.*, the Matthew effect, describing phenomenon rich get richer [15]) affects changes on the groups level. However before answering this question the number of extensive researches on the evolution of groups has to be conducted in order to deepen the knowledge in this field.

Although some methods concerning prediction of some aspects (e.g., determining lifespan) of group evolution have been proposed so far, the task of predicting future evolution event for a group was not studied so extensively, except for the methods presented in [16–18]. Methods described in [16,17] focus on predicting next event for a group based on its previous states (called also profiles). A group profile is described by a set of metrics (explained later in detail in Section 5.3) and prediction is conducted by the classifier. However, only two kinds of features: a group size and transition type were utilized in [16,17] for reasoning. To some extent a similar approach is proposed in [18], where the authors also consider group profiles, but they introduced more characteristics describing groups such as properties of influential members in a group and topics discussed by a group. They assumed that group events are independent (they use classifiers to predict absence or presence of a single event, *i.e.*, binary prediction), but they did not investigate history of group states (only the previous change). In this paper, we concentrate on the in-depth analysis of reasoning process, in particular: (1) the evaluation of the influence of the number of previous group states (length of history) on quality of evolution prediction as well as (2) the selection of proper input features describing groups from the pool of diverse structural measures (31 for each time frame), including aggregated ones.

## 4. General Concept

Prediction of social community evolution consists of four main phases depicted in Figure 1:

(1)  collecting data and its splitting into time frames (Figure 1a);
(2)  extraction of social networks for each period and social community identification (Figure 1b);
(3)  detection of changes (events) in groups for the following time windows and identification of chains preceding the recent state of the group (Figure 1c);
(4)  building the predictive model (learning the classifier) and its validation (Figures 1d and 2).

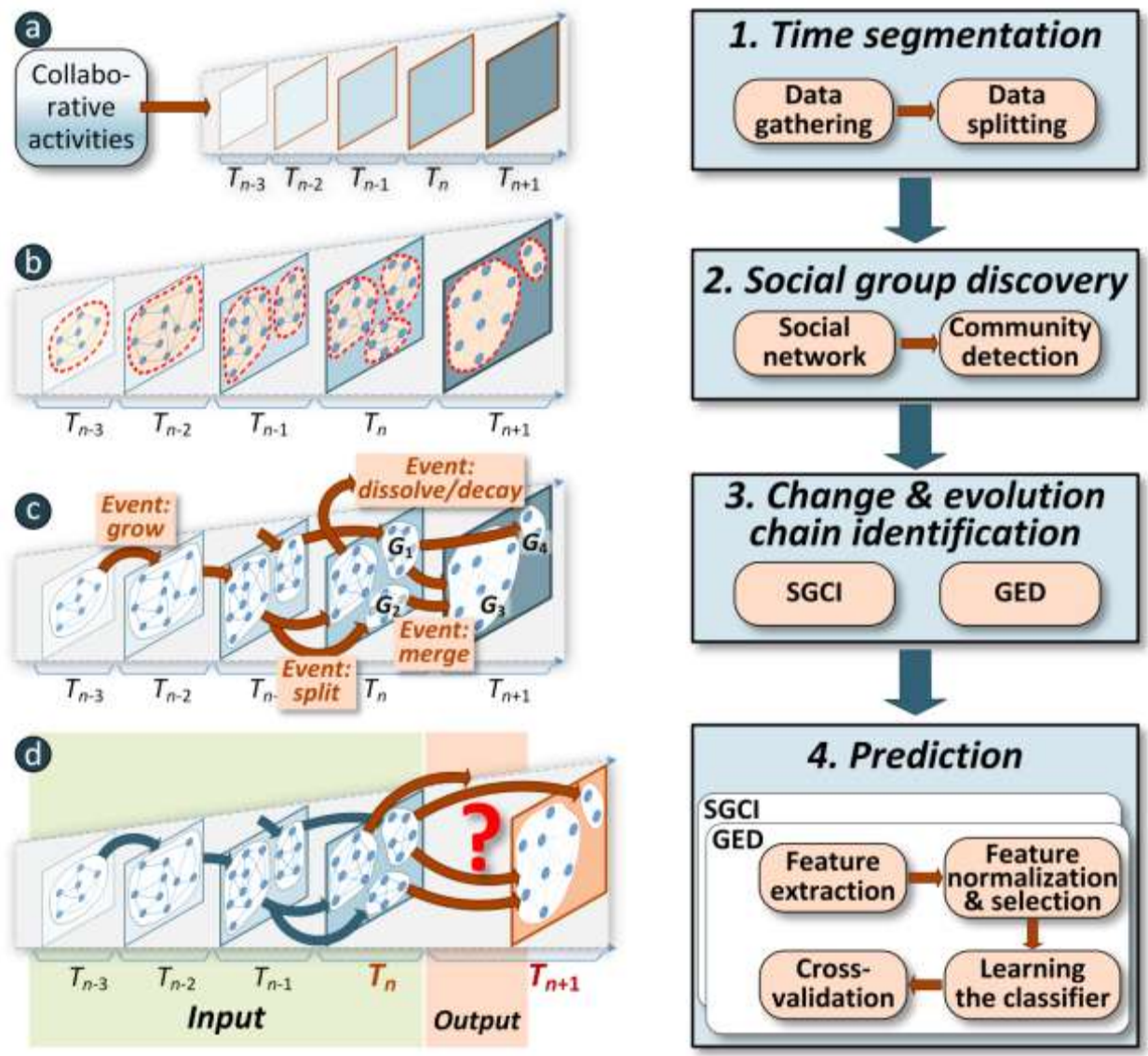

**Figure 1.** The four main phases in prediction of social community evolution.

First, some source data about common activities, collaborative work or mutual communication needs to be gathered from IT services or databases over reasonable long time to capture evolution in human behaviour and mutual relationships (Figure 1a). In the experimental part, three data sources were exploited: posts and discussions in blogosphere, co-authorship of scientific papers and Facebook. All of them correspond to different human activities over dozens of months with various engagement and variability in interpersonal contacts.

In the next stage (Figure 1b), a separate social network needs to be created from the source data separately for each time frame. In general, this network can form either a weighted or unweighted graph, either directed or undirected one. Weights in the weighted case reflect the intensity level of common activity. In the experimental studies (see Section 7) only weighted graphs were exploited even though their weights were used for prediction only (at evaluation of descriptive structural features)—not for community detection. Two of the graph time series were directed whereas one was undirected.

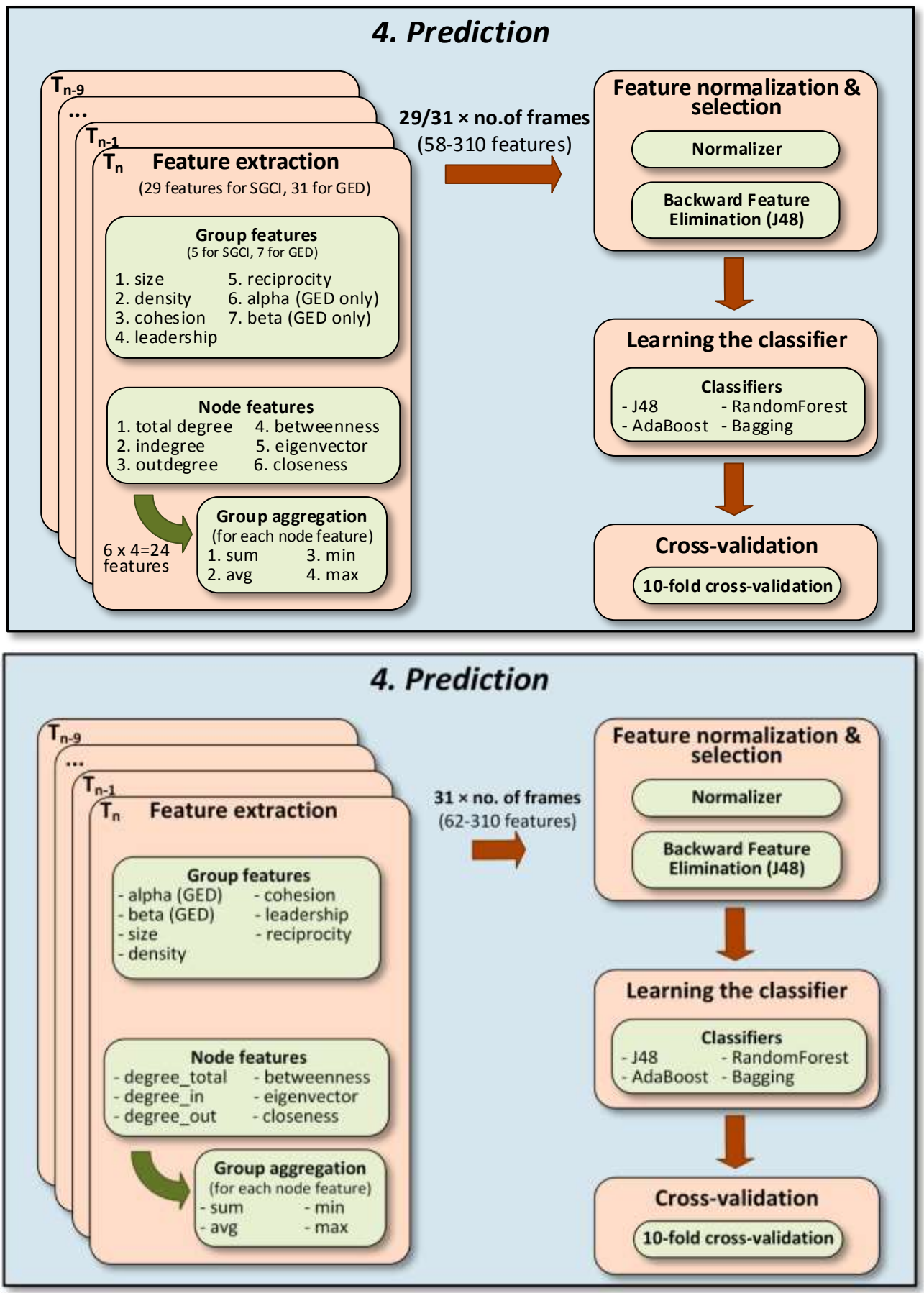

**Figure 2.** The last phase in prediction of social community evolution.

Having the network, social community are discovered by means of any clustering method. In this research, the clique percolation method (CPM) [19,20] was used for extraction of overlapping groups, although one could utilize any other algorithm.

The crucial goal of the third phase is to detect changes of social communities between following time frames: $T_{n-1} \rightarrow T_n$, *i.e.*, events like group *splitting*, *growing*, *merging*, *dissolving* and so on (Figure 1c). Two different algorithms developed by the authors were independently utilized for that purpose, see Section 5 for details:

(1)  Stable Group Changes Identification Algorithm (SGCI),
(2)  Group Evolution Discovery Algorithm (GED).

Based on the changes identified, an evolution chain can be created for each group $G_i$ from $T_n$. Such chain consists of all other preceding groups from the previous time frames ($T_{n-1}$, $T_{n-2}$, $T_{n-3}$, *etc.*) the recent group $G_i$ comes from. Additional information (metrics described in section 5.3) about related changes that formed group $G_i$ is added. Overall, it may happen that a group has been created from two or more other groups—through *merging*, e.g., group $G_3$ came into being from $G_1$ and $G_2$. In such case, two separate evolution chains are being established for $G_3$, one with group $G_1$ and one with $G_2$. It could be even multiplied by more *merging* events in the following time frames. In general, many evolution chains may be assigned to one group.

The last, fourth stage is the prediction by means of machine learning methods (Figures 1d and 2). It has been performed independently for SGCI and GED algorithms to enable their comparison. For each group $G_i$ in time frame $T_n$, a set of descriptive, mainly structural features is computed (see Section 5.3). These features correspond to both the state of group $G_i$ within the social network in $T_n$ (29 features for *SGCI* and 2 more for *GED*), as well as its identified ancestors (previous groups) and transitions in the preceding periods $T_{n-1} \rightarrow T_n$, $T_{n-2} \rightarrow T_{n-1}$, $T_{n-3} \rightarrow T_{n-2}$, …, *etc.* This calculation is based on the items coming from the group evolution chain detected in the third phase. Through this transformation of evolution chains into descriptive features (attributes, variables), we obtain even hundreds of features reflecting historical evolution of each group until $T_n$. For each period a separate feature set is calculated. In total, we obtain (*no. of features for a period*) × (*no. of periods*) features for each considered group in $T_n$. Groups with their descriptive features from each time frame $T_n$ (except the first periods that possess too short history) are put into one set of group instances ready to build predictive model. Since there may be several evolution chains for a given group $G_i$, each of them corresponds to another case. As a result, group $G_i$ occurs as many times in the learning set as many evolution chains were detected for it.

To check usability of the features, an additional feature selection method is applied to remove unnecessary attributes and the ones that potentially provide too many awkward disturbances. Additionally, to enable application of various classification models, that may have some limitations related to feature domains, the normalization process is performed for each feature. It is carried out by linear transformation into the range [0;1] and min-max approach, *i.e.*, the minimal value is moved to 0 and maximal to 1.

The groups from all periods (excluding the first ones) with their features (input) together with information about their following transition in $T_{n+1}$ (output) are used to learn the classifier, *i.e.*, build a classification model. For the validation purpose (to evaluate a quality of prediction), the entire set of these cases (chains of groups) is randomly split into 10 partitions to enable 10-fold cross-validation: learning on nine sets, testing on the remaining 10th and repetition this process 10 times for another remaining testing set.

Furthermore, to analyse the evolution chain profiles, independent classification models were built for various lengths of historical data (evolution chain) used to describe the group, *i.e.*, we considered two, three, four, …, *etc.* previous changes of the given group.

In general, a group may be involved in many events for a given transition $T_n{\rightarrow}T_{n+1}$. For example, group $G_1$ in Figure 1c was involved in both *merging* with $G_2$ into $G_3$ and *splitting* into $G_3$ and $G_4$. In this study, we used only typical multi-class classification method; its output is one out of many classes from the fixed set of event types. Hence, the learned model is able to predict only one future event for a given group described by the assigned features derived from one evolution chain. To enable multiple output (many possible events) another solution—multi-label classification would need to be applied [21,22]. This is, however, much more complex and is rather a matter of further research.

## 5. Predicting Group Evolution in the Social Network

In this chapter two approaches to group evolution prediction in social network are presented. Both approaches are similar but the first one is based on the SGCI method results and the second is based on the GED method results. The SGCI method was introduced in [24] (the concept of stable groups was used from the method described in [25]). In [26] authors of SGCI described a tool for visualisation of group evolution based on the SGCI method. A detailed explanation of the GED method can be found in [23], and in [24] both methods of identification events in group evolution are compared.

### 5.1. Predicting Group Evolution Using SGCI Results and the Notion of Dominating Event

One of the algorithms used in presented research, which identifies groups of interacting users is the SGCI algorithm [24,25]. The main idea of the algorithm is finding groups in subsequent time frames and identification of their continuation. A particular feature of the algorithm concerns the identification of groups which fulfil the stability condition—their duration is longer than a given number of time.

Different kinds of events were identified, which describe a method of transformation of stable groups between considered time frames and priorities of these events were defined as follows (starting from the highest one): *constancy*, *change size*, *split*, *merge*, *addition*, *deletion*, *split_merge*, *decay*. The reasons behind such choice of the order of importance of events were described in [17].

The chains of the following states of stable groups, with given length, are considered. The state is represented by values of the selected measures, described in Section 5.3. The applied prediction algorithm is based on finding next type of events transforming groups, taking into consideration their previous states in analysed chains. For each group in a given time frame there is a possibility to predict many events (which arise from different considered chains from the past), so a notion of a dominating event, defined on the basis of event priorities, was introduced. The reason for selected order is that some events, such as addition or deletion mean small change for groups. Moreover, some events cannot coexist with other ones (described in [17]) and position in order of such events is meaningless (such as the *decay* event).

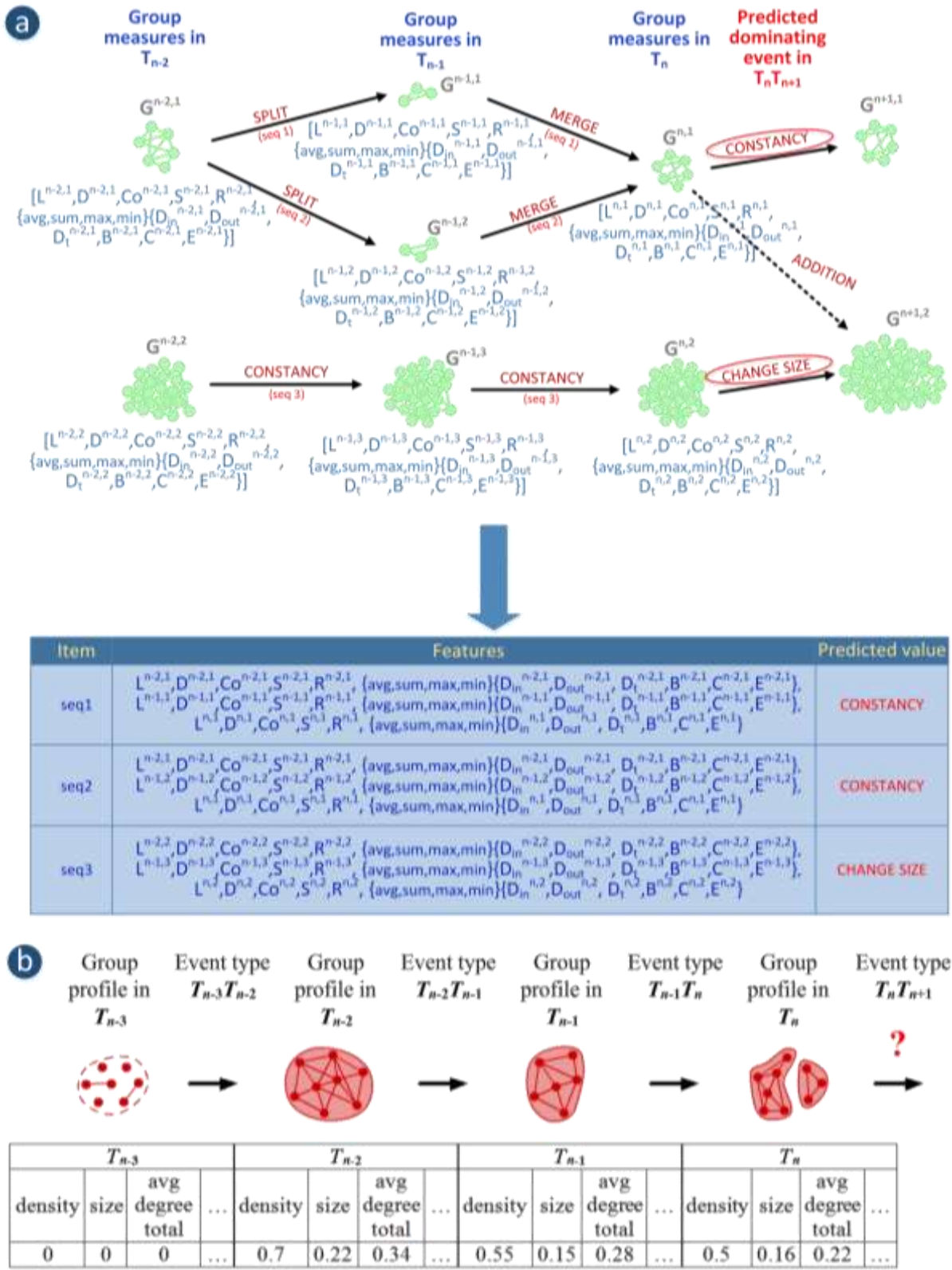

**Figure 3.** Two approaches to group evolution prediction: (**a**) using the SGCI method—example with sequences of group measures from 3 time frames (1 present group state and 2 earlier group states) and predicted dominating event; (**b**) using the GED method—the sequence of events for a single group together with its profiles as well as its target class-event type in $T_nT_{n+1}$ (the chain corresponds to one case in classification).

Figure 3a explains the main idea of the presented algorithm and the notion of the dominating events. Three sequences (labelled as *seq1*, *seq2* and *seq3*) of the group states, each of them with the length equal two, are presented. Each state is described by a vector of measures. For instance the state of the group $G_1$ in time $T_n$, called $G^{n,1}$, is expressed by a following vector: $[L^{n,1}, D^{n,1}, Co^{n,1}, S^{n,1}, R^{n,1},$ $\{avg, sum, max, min\}\{D_{in}^{n,1}, D_{out}^{n,1}, D_t^{n,1}, B^{n,1}, C^{n,1}, E^{n,1}\}]$, where upper index of each measure concerns the number of time frame ($n$) and the number of the group (1). The values of functions (avg, sum, max, min) are calculated for values of given measure ($D_{in}^{n,1}, D_{out}^{n,1}, D_t^{n,1}, B^{n,1}, C^{n,1}, E^{n,1}$) of all members of considered groups (see Section 5.3).

In Figure 3a *seq1* represents a sequence of states of groups $G_{n-2,1}$, $G_{n-1,1}$ and $G_{n,1}$ and we want to predict the next evolution event for sequence *seq1* and group $G_{n,1}$. As we can see in Figure 3a, this group has two events assigned: *constancy* (transition between $G_{n,1}$ and $G_{n+1,1}$) and *addition* (transition between $G_{n,1}$ and $G_{n+1,2}$). According to the introduced concept of dominating events and chosen priorities of events, the predicted dominating event is the *constancy*. The table in Figure 3a summarizes the types of predicted events for each considered sequence.

*5.2. Predicting Group Evolution Using GED Results*

The idea of using the GED method results [23] to predict group evolution was presented in [16]. The initial idea was to use a simple sequence, which consists of groups size and events between consecutive timeframes as an input for the classifier. The learnt model should be able to produce very good results even for simple classifiers. The initial results were quite encouraging especially for tree classifiers, thus the next step was to transform the initial concept into a method.

Firstly, instead of simple group size, the entire group profile, consisting of dozens of metrics and their aggregations (centrality measures were aggregated on the level of a group-for details please see Section 5.3) was built. This profile describes the state of group in selected timeframe before and after particular transition (event).

Secondly, instead of constant sequence length, the different chain lengths (sequence lengths) were introduced, to find out what would be the results if we use shorter/longer sequences (more preceding events and group measures).

As an example, the 4-step sequence is used (Figure 3b). Obviously, the event types vary depending on the individual groups, but the time frame numbers were fixed, for one method execution. It means that for each predicted event, four group profiles in four previous time frames together with three associated events are identified as the input for the classification model, separately for each group. A single group in a given time frame ($T_n$) is a case (instance) for classification, for which its event $T_nT_{n+1}$ is being predicted.

The sequence presented in Figure 3b is used as an input for classification. The first part of the sequence is used as input features (variables), *i.e.*, (1) Group profile in $T_{n-3}$; (2) Event type $T_{n-3}T_{n-2}$; (3) Group profile in $T_{n-2}$; (4) Event type $T_{n-2}T_{n-1}$; (5) Group profile in $T_{n-1}$; (6) Event type $T_{n-1}T_n$; (7) Group profile in $T_n$. A predictive variable is the next event for a given group. Thus, the goal of classification is to predict (classify) Event $T_nT_{n+1}$ type–out of the six possible classes: *i.e.*, (1) *growing*, (2) *continuing*; (3) *shrinking*; (4) *dissolving*; (5) *merging* and (6) *splitting*. The *forming* event was excluded since it can only start the sequence.

Main differences between both methods to predict groups evolutions include:

- usage of different methods of groups evolution (SGCI and GED, respectively)
- the concept of dominating event in approach using SGCI method
- usage of additional, specific measures for prediction of events in approach using GED method (metrics alpha and beta which are utilized internally in the process of determining groups transitions in consecutive time frames in GED method)
- different generation of chains for *split/splitting* event (for GED if the last group in a chain has assigned *splitting* events with multiple groups in the next time frame, then for each *splitting* transition for the considered group the identical chain is generated, but with SGCI only one such chain is generated).

### 5.3. Measures Used To Describe Group Profile

There are up to 31 features extracted for each group and for each period, see Figure 2. They describe the group itself like its size, average density, cohesion, leadership, reciprocity and additionally for the GED method: alpha and beta. There are five such features for SGCI and seven for GED.

Independently, the aggregated structural features for nodes that belong to a given group are computed. The following six measures were used for that purpose: node total degree, indegree, outdegree, betweenness, eigenvector and closeness. For each of these six features some aggregations over all nodes in the group are calculated, namely: *sum*, *average*, *minimum* and *maximum* value. It means that in case of *total degree* and *sum*, the sum of total degree for all nodes in the group is computed. For a single measure four aggregations are created, *i.e.*, for *total degree* we have: *avg_total_indegree*, *sum_total_indegree*, *min_total_indegree* and ***max_total_indegree***. As a result, we obtain $6 \times 4 = 24$ aggregated features for each group.

Together with the group features, it makes $5 + 24 = 29$ features for *SGCI* and 31 features for GED. Their values are independently computed for each period in the evolution chain, e.g., for 5-period chain we have $29 \times 5 = 145$ features (*SGCI*) or 155 for GED, for each single group in $T_n$. The explanations for the features/structural measures are presented below:

- **group size**–the number of nodes in the group,
- **density**–a measure expressing how many connections between nodes are present in the group in relation to all possible connections between them [1]:

$$D = \frac{\sum_i \sum_j a(i,j)}{n(n-1)} \quad (4)$$

where function $a(i,j)$ has value 1 when there is connection from node $i$ to node $j$,

- **cohesion**–a measure characterising strength of connections inside the group in relation to the connections outside the group (incident with the group members) [1]:

$$C = \frac{\dfrac{\sum_{i \in G} \sum_{j \in G} w(i,j)}{n(n-1)}}{\dfrac{\sum_{i \in G} \sum_{j \notin G} w(i,j)}{N(N-n)}} \quad (5)$$

where $w$ is a function assigning the weight between nodes, $G$ is a group, $n$ is the number of nodes in the group and $N$ is the number of nodes in the entire network,

- **leadership**–a measure describing centralization in the graph or group (the largest value is for a star network) [27]:

$$L = \sum_{i=1}^{n} \frac{d_{max} - d_i}{(n-2)(n-1)} \qquad (6)$$

where $d_{max}$ means the maximum value of degree in the group and $n$–the number of nodes in the group,

- **reciprocity**–a fraction of edges that are reciprocated [28]:

$$R = \frac{1}{m} \sum_{ij} a(i,j)a(j,i) \qquad (7)$$

where $m$ is the total number of edges in the network and function $a(i,j)$ has value 1 if there is a connection from node $i$ to node $j$,

- **alpha**–the GED inclusion measure of group $G_i$ from time frame $T_n$ in group $G_j$ from time frame $T_{n+1}$ [23] (a measure used only in approach utilizing the GED method),
- **beta**–the GED inclusion measure of group $G_j$ from time frame $T_{n+1}$ in group $G_i$ from time frame $T_n$ [23] (a measure used only in approach utilizing the GED method),
- **indegree**–a node measure defining the number of connections directed to the node [27]:

$$D_{in} = \sum_{i} a(j,i) \qquad (8)$$

where function $a(j,i)$ has value 1 if there is a connection from node $j$ to node $i$,

- **outdegree**–a node measure determining the number of connections outgoing from the node [27]:

$$D_{out} = \sum_{i} a(i,j) \qquad (9)$$

where function $a(i,j)$ has value 1 if there is a connection from node $i$ to $j$,

- **total degree**–sum of indegree and outdegree:

$$D = D_{in} + D_{out} \qquad (10)$$

- **betweenness**–a node measure describing the number of the shortest paths from all nodes to all others that pass through that node [27]:

$$B = \sum_{i \neq j \neq v} \frac{\sigma_{ij}(v)}{\sigma_{ij}} \qquad (11)$$

where $\sigma_{ij}(v)$ align is the total number of the shortest paths from node $i$ to $j$ and $\sigma_{ij}(v)$ is the number of those paths that pass through $v$,

- **closeness**–a node measure defined as the inverse of the farness, which in turn, is the sum of distances to all other nodes [27]:

$$C = \sum_{i \neq j} \frac{1}{d(i,j)} \tag{12}$$

where function $d(i,j)$ is distance from node $i$ to $j$,

- **eigenvector**–a node measure indicating the influence of a node in the network [29].

## 6. Dataset and Experiment Setup

### 6.1. Dataset Description

Experiments were conducted on three different datasets. The first dataset is the DBLP network dataset which contains the undirected collaboration graph of authors of scientific papers. This dataset is publicly available [30] and the graph has 1,248,427 vertices and 17,631,144 edges. For the analysis we used data range from 1990 to 2009—this period of time was divided into 20 disjoint time frames lasting 1 year each.

The second one is the Facebook dataset. It is also publicly available [31] and comprises directed network of posts to other user's wall on Facebook. The network contains 46,952 vertices and 876,993 edges. The analysis was conducted on the data from range 8 January 2015–2 November 2015which was split into 42 overlapping time frames, each lasting 60 days and each overlaps 50% with the neighbouring one.

The last one is the Salon24 dataset which contains data from the www.salon24.pl portal consisting of blogs from different domains, most of them are political ones. The data consists of 26,722 users, 285,532 posts and 4,173,457 comments. Tests were conducted on the data from range 4 April 2010–31 March 2012. The analysed period of time was divided into time frames, each lasting 7 days and neighbouring time frames overlap each other by 4 days. In this period of time there are 182 time frames.

### 6.2. Group Extraction

After dividing data on time frames, the next step was discovering groups in each of them using. CPM method (CPMd version) from CFinder tool (http://www.cfinder.org/). CPM requires k parameter which decides about minimum size of groups. We set values for *k* equal to 3, 7 and 5 for the Facebook, DBLP and Salon24 dataset, respectively. Different values of this parameter were motivated by performance issues.

### 6.3. Group Sizes

Each dataset has different characteristics of groups (presented in Figure 4). Most medium size groups belong to the DBLP dataset (with 7–50 members). The smallest groups are from Facebook dataset. Contrary, Salon24 dataset contains the biggest groups. These differences could be explained by the different nature of these datasets. The collaboration network (DBLP dataset) and writing posts to other users' wall in a social network service (Facebook dataset) cause the formed groups to not be large.

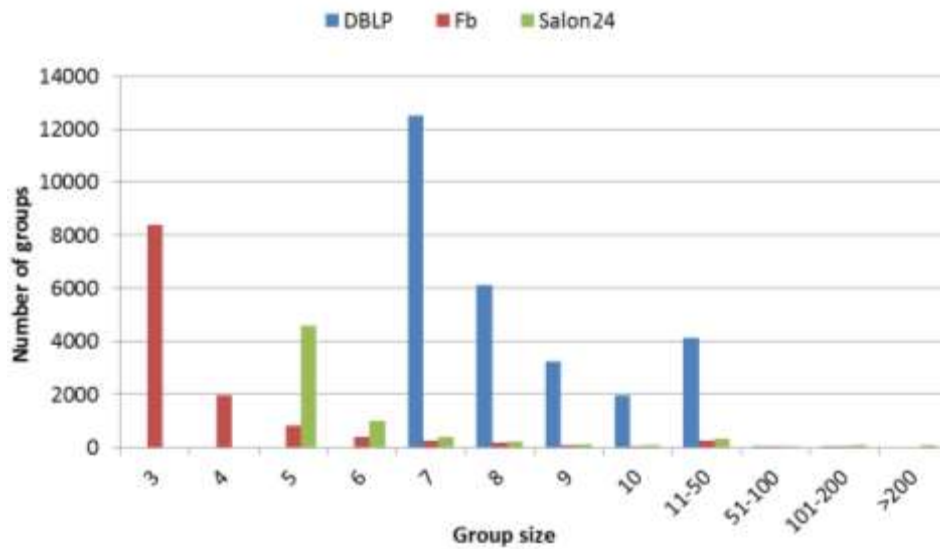

**Figure 4.** The number of groups with a given size.

*6.4. Experiment Setup*

The experiments using the SGCI method were conducted using the following parameters (described in detail in [17]): $ds$ = 50 (except for the Salon24 dataset, where this parameter had a value equal to 25), $sh$ = 10 and $dh$ = 0.05. The value of parameter $MJ$ was different for each dataset: for DBLP $MJ$ = 0.4, for Facebook–0.5 and for Salon24–0.65. Different values of parameters were motivated both by performance issues and need to obtain sufficient number of events during groups evolution.

The GED method was run on the datasets with all combinations of GED parameters [23] from the set {50%, 60%, 70%, 80%, 90%, 100%}. The social position measure [32] (measure similar to weighted page rank) was utilized as the node importance measure. The Facebook and DBLP datasets needed additional run with parameters equal to 30%.

To describe the group profile at specific time frame following measures were used: size, density, cohesion, leadership, reciprocity, avg_indegree, sum_indegree, min_indegree, max_indegree, avg_outdegree, sum_outdegree, min_outdegree, max_outdegree, avg_total_degree, sum_total_degree, min_total_degree, max_total degree, avg_betweenness, sum_betweenness, min_betweenness, max_betweenness, avg_closeness, sum_closeness, min_closeness, max_closeness, avg_eigenvector, sum_eigenvector, min_eigenvector, max_eigenvector. Additionally, both inclusions (alpha, beta measures) were used with the GED method.

The experiment was executed in KNIME (www.knime.org) with Weka plugin. Four different classifiers were utilized with default settings (Table 1). To evaluate the quality of the methods, 10-fold cross-validation was used [33] with stratified sampling separately for the GED and SGCI results. The measure selected for presentation and analysis of the results was F-measure, which is the harmonic mean of precision and recall.

**Table 1.** Classifiers used.

| Short Name | Name |
|---|---|
| J48-C4.5 decision tree | C4.5 decision tree [34] |
| RandomForest | Random forest [35] |
| AdaBoost(J48) | Adaptive Boosting [36] |
| Bagging(REPTree) | Bootstrap aggregating [37] |

All classifiers were utilized for both approaches (SGCI and GED) and their results are presented below.

## 7. Experiments

### 7.1. Predicting Group Evolution Using the SGCI Results

A distribution of events in all three considered datasets is presented in Figure 5. In the DBLP and Facebook datasets, change size is the most frequent event. Salon24 has a significantly different characteristic, it possesses much more events leading to several groups reorganisation (*addition*, *deletion*, *merge* and *split*).

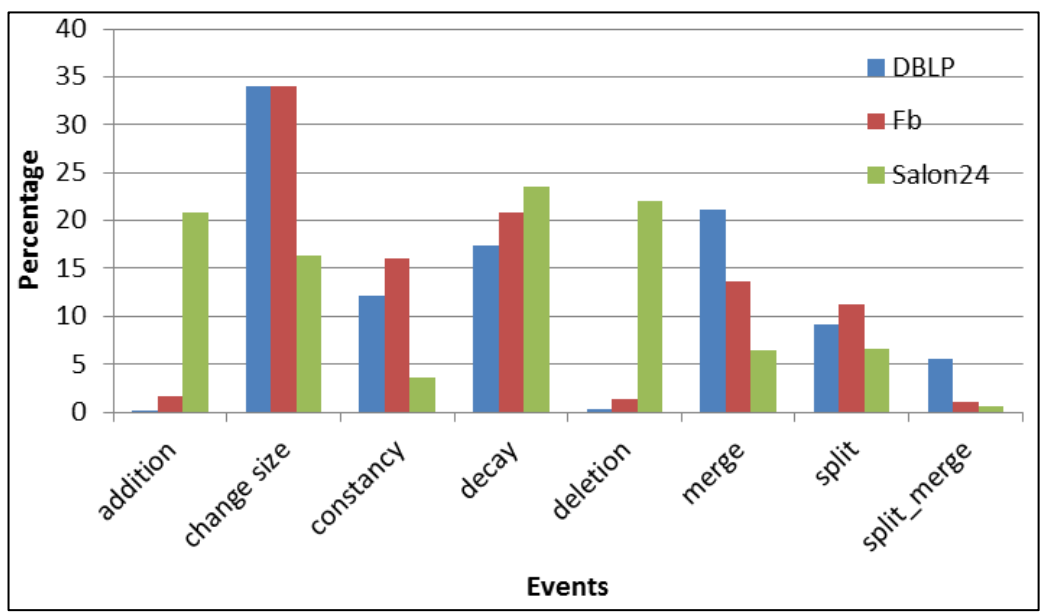

**Figure 5.** SGCI: distribution of the event types in all datasets.

In Table 2 there are overall numbers of chains identified in each analysed dataset, while detailed statistics regarding frequencies of types of events in datasets are presented in Tables 4–6. As one can see in Table 2 the overall number of chains in DBLP and Facebook decreases while the length of chain increases. In Salon24.pl the trend is inversed and, the number of chains decreases with the length of chains. This behaviour stems from the fact that the Salon24 dataset contains much more events that increase the number of chains when we consider longer chains, *i.e.*, events such as *deletion*, *split* and *spit_merge* together constitute significantly higher part of all events in comparison with other datasets.

**Table 2.** SGCI: the number of evolution chains for particular chain length.

| Chain Length | DBLP | Facebook | Salon24 |
|:---:|:---:|:---:|:---:|
| 2 | 2980 | 3027 | 2119 |
| 3 | 2581 | 2759 | 5999 |
| 4 | 2051 | 2094 | 5005 |
| 5 | 1919 | 1831 | 10,712 |
| 6 | 1754 | 1575 | 9895 |
| 7 | 1120 | 1401 | 15,076 |
| 8 | 744 | 1314 | 18,735 |
| 9 | 603 | 1280 | 29,690 |
| 10 | 417 | 1141 | - |

**Table 3.** SGCI: the number of evolution chains for particular event type and particular chain length in the DBLP dataset.

| Chain Length | Addition | Change Size | Constancy | Decay | Deletion | Merge | Split |
|:---:|:---:|:---:|:---:|:---:|:---:|:---:|:---:|
| 2 | 7 | 981 | 340 | 846 | 5 | 471 | 330 |
| 3 | 7 | 569 | 166 | 964 | 3 | 520 | 352 |
| 4 | 4 | 431 | 126 | 548 | 3 | 516 | 423 |
| 5 | 3 | 432 | 106 | 379 | 1 | 499 | 499 |
| 6 | 0 | 428 | 82 | 296 | 1 | 532 | 415 |
| 7 | 0 | 334 | 72 | 135 | 0 | 381 | 198 |
| 8 | 0 | 219 | 39 | 146 | 0 | 229 | 111 |
| 9 | 0 | 182 | 29 | 99 | 0 | 178 | 115 |
| 10 | 0 | 106 | 16 | 82 | 0 | 135 | 78 |

**Table 4.** SGCI: the number of evolution chains for particular event type and particular chain length in the Facebook dataset.

| Chain Length | Addition | Change Size | Constancy | Decay | Deletion | Merge | Split |
|:---:|:---:|:---:|:---:|:---:|:---:|:---:|:---:|
| 2 | 23 | 1137 | 416 | 840 | 32 | 298 | 281 |
| 3 | 17 | 854 | 286 | 1078 | 18 | 295 | 211 |
| 4 | 23 | 680 | 202 | 714 | 20 | 247 | 208 |
| 5 | 8 | 623 | 160 | 624 | 23 | 204 | 189 |
| 6 | 11 | 541 | 134 | 499 | 11 | 215 | 164 |
| 7 | 13 | 457 | 139 | 425 | 11 | 195 | 161 |
| 8 | 14 | 434 | 118 | 389 | 11 | 170 | 178 |
| 9 | 9 | 394 | 99 | 438 | 19 | 168 | 153 |
| 10 | 5 | 324 | 87 | 426 | 16 | 147 | 136 |

**Table 5.** SGCI: the number of evolution chains for particular event type and particular chain length in the Salon24 dataset.

| Chain Length | Addition | Change Size | Constancy | Decay | Deletion | Merge | Split |
|---|---|---|---|---|---|---|---|
| 2 | 185 | 615 | 72 | 683 | 255 | 125 | 184 |
| 3 | 920 | 764 | 102 | 3638 | 157 | 216 | 202 |
| 4 | 603 | 1098 | 68 | 2280 | 444 | 214 | 298 |
| 5 | 1334 | 1510 | 104 | 6773 | 340 | 338 | 313 |
| 6 | 1064 | 2170 | 138 | 5201 | 398 | 464 | 460 |
| 7 | 1860 | 2573 | 158 | 8597 | 594 | 563 | 731 |
| 8 | 2065 | 3357 | 365 | 9917 | 912 | 920 | 1199 |
| 9 | 4126 | 3900 | 533 | 16,498 | 1151 | 1875 | 1607 |

In Figures 7–11, 13–19 and 21–27 F-measures for each dataset and each event are presented. For each datasets the RandomForest, J48-based AdaBoost, Bagging classifiers are used, for Facebook datasets, (Figures 21–27) Feature Selection classifier is also calculated.

### 7.1.1. DBLP Dataset

In Figure 6 and Table 3 one can see a distribution and numbers of each kind of events for given chain lengths. The number of events decreases with the increase in the considered chain length. The number of events is low in comparison to other datasets, especially *addition* and *deletion* events, which do not appear in 6-length or longer chains (thus we do not assess quality of prediction for them in this dataset).

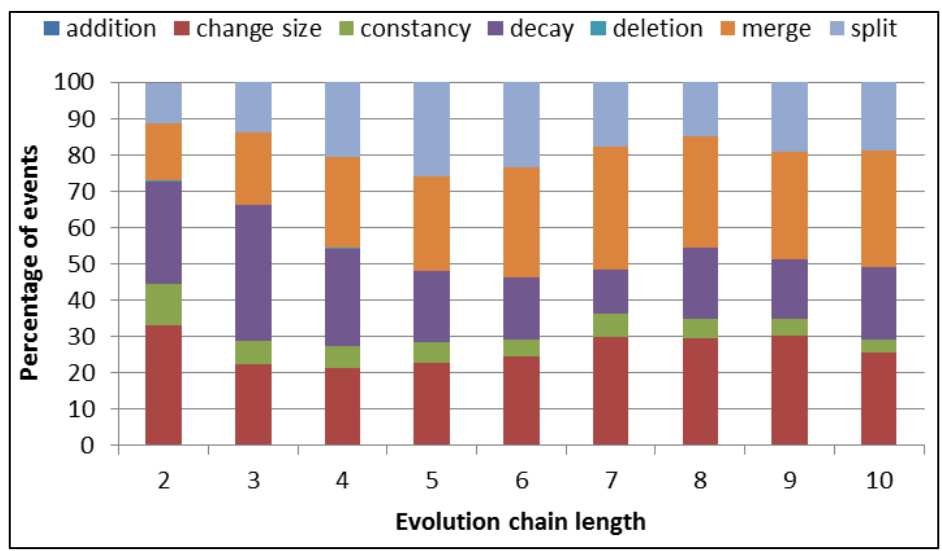

**Figure 6.** SGCI: distribution of the event types for events being predicted in the DBLP dataset.

In Figures 7–11 F-measures for DBLP depending on chain length for each event are presented. For each case, the best accuracy is obtained for RandomForest and AdaBoost classifiers. The results differ regarding initial F-measure value for shortest chain length (equal 2) and the chain length value for which the maximum F-measure value is obtained. The highest accuracy is obtained for most frequent events, such as *change-size* and *decay*.

Generally, F-measure achieves very high quality for relatively short chains values (equal 5 or 6). All events are well classified (especially in longer chains)-only *constancy* event obtained a little worse results in short chains with comparison to other events. Please also note that *constancy* events are not as frequent as other ones.

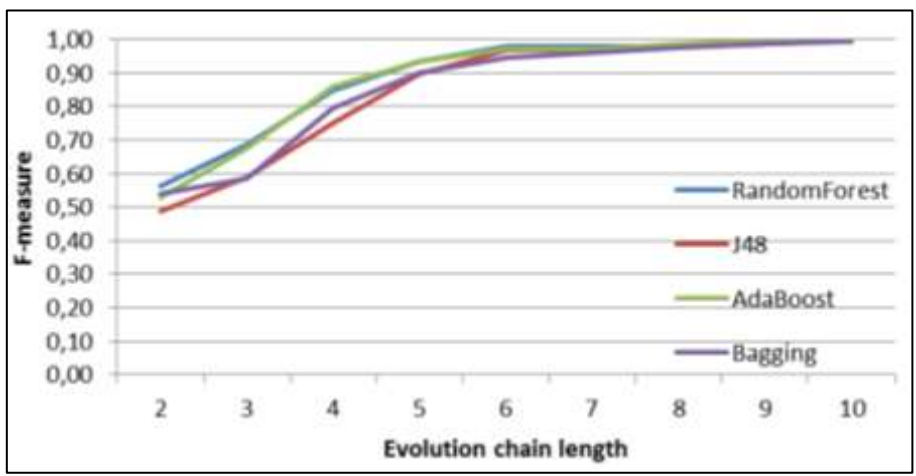

**Figure 7.** SGCI: results of event classification for *change size* event in the DBLP dataset.

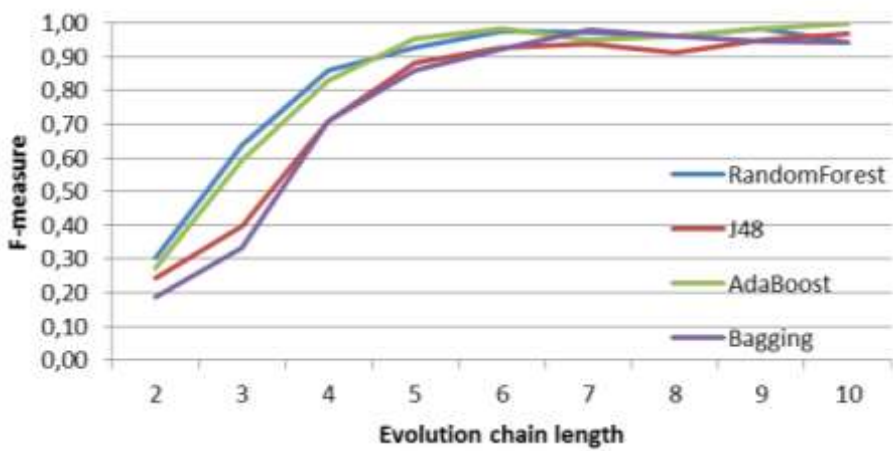

**Figure 8.** SGCI: results of event classification for *constancy* event in the DBLP dataset.

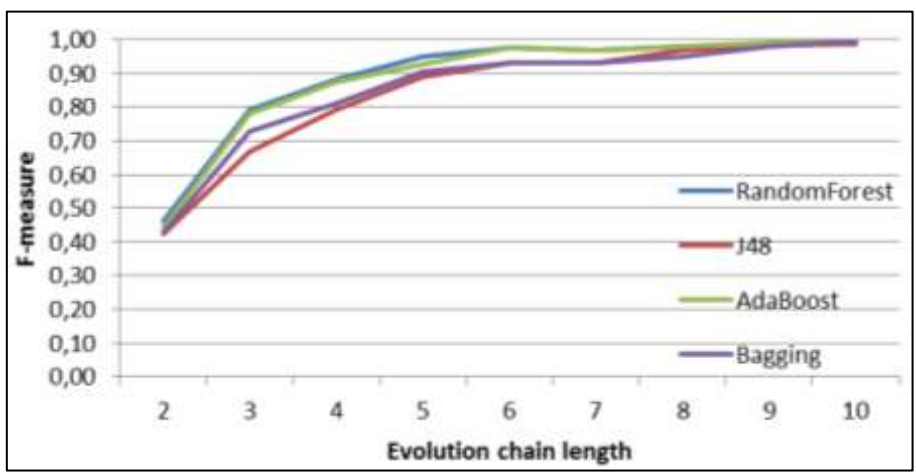

**Figure 9.** SGCI: results of event classification for *decay* event in the DBLP dataset.

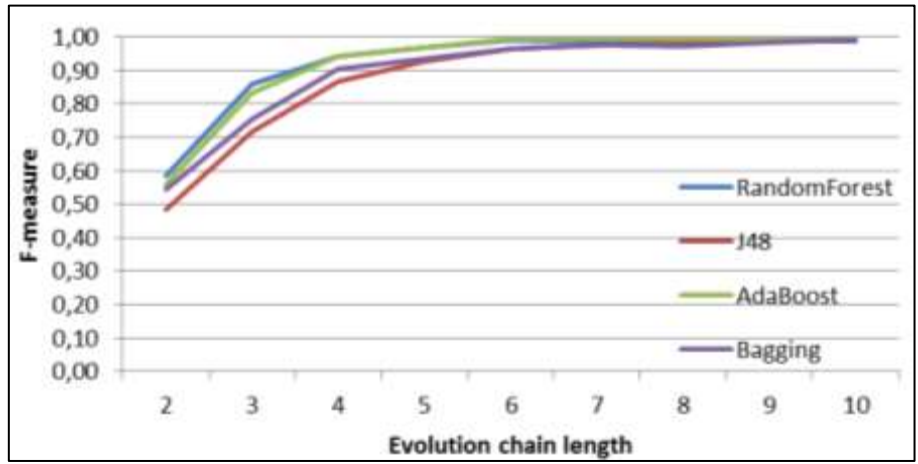

**Figure 10.** SGCI: results of event classification for *merge* event in the DBLP dataset.

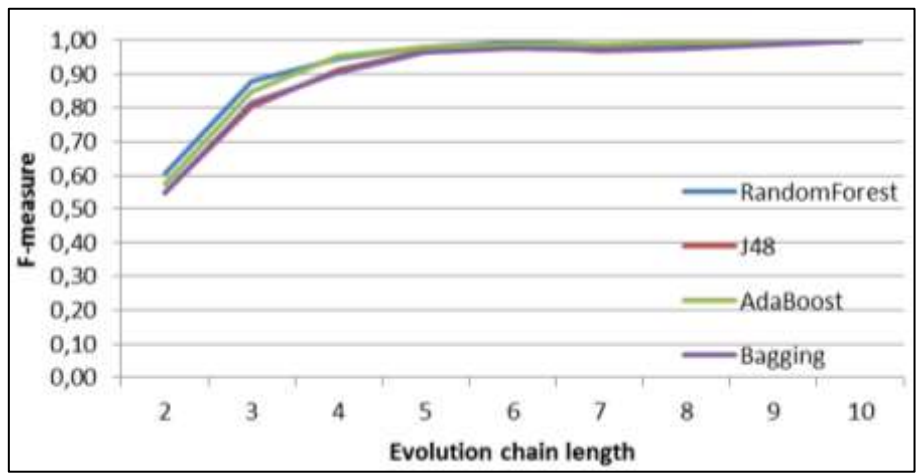

**Figure 11.** SGCI: results of event classification for *split* event in the DBLP dataset.

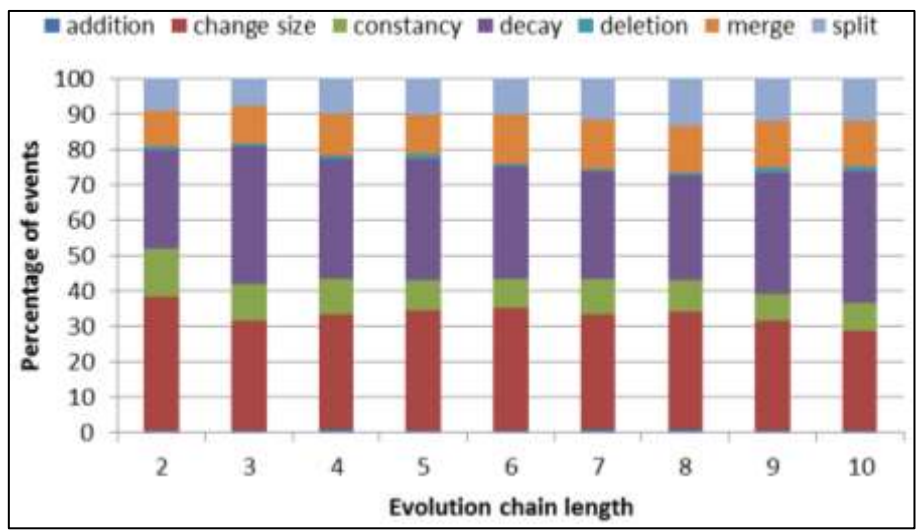

**Figure 12.** SGCI: distribution of the event types for events being predicted in the Facebook dataset.

7.1.2. Facebook Dataset

The Facebook dataset has different a profile than the DBLP one. It contained much more groups, especially small ones and much more events. It is closely related with different nature of social media, especially their higher dynamics. The changes in group structures take place there faster and are more frequent. The characteristic of changes of F-measure is also different. Like the DBLP dataset, the number of events decreases with the increase in the considered chain length.

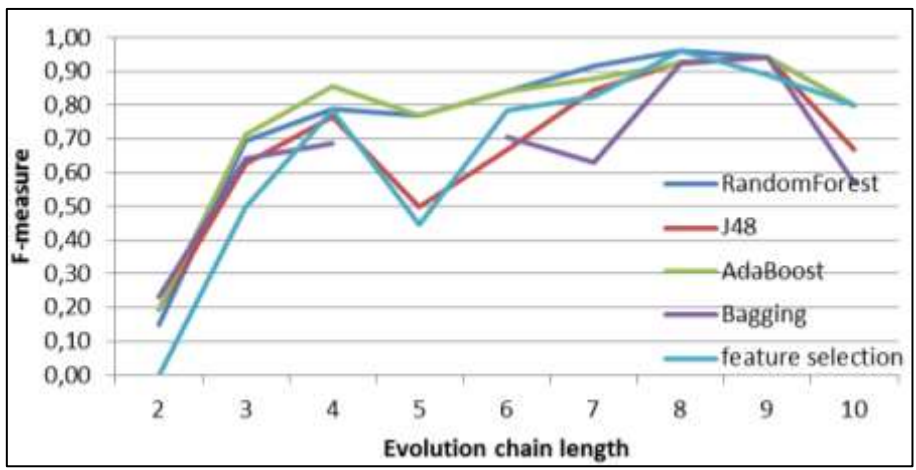

**Figure 13.** SGCI: results of event classification for *addition* event in the Facebook dataset.

Values of F-measure for the considered cases (Figures 13–19) have the shape different to the one for the DBLP dataset. Again, RandomForest and AdaBoost are the best classifiers. The highest differences between these two classifiers and J48 or Bagging are for *addition* (Figure 13) and *deletion* (Figure 17). It is connected with a relatively low numbers of these events. The best results (considering F-measure value of the shortest chain length and chain-length for which F-measure values are close to the maximum value) are obtained for *split* and *change size* events.

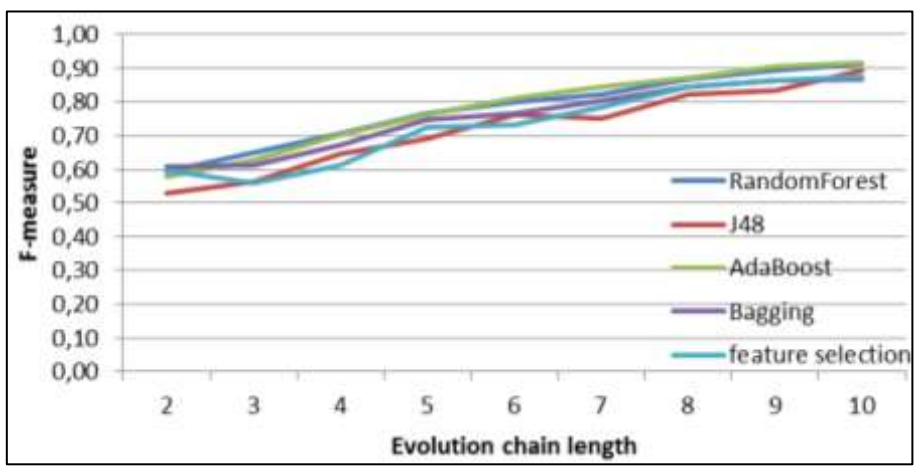

**Figure 14.** SGCI: results of event classification for *change size* event in the Facebook dataset.

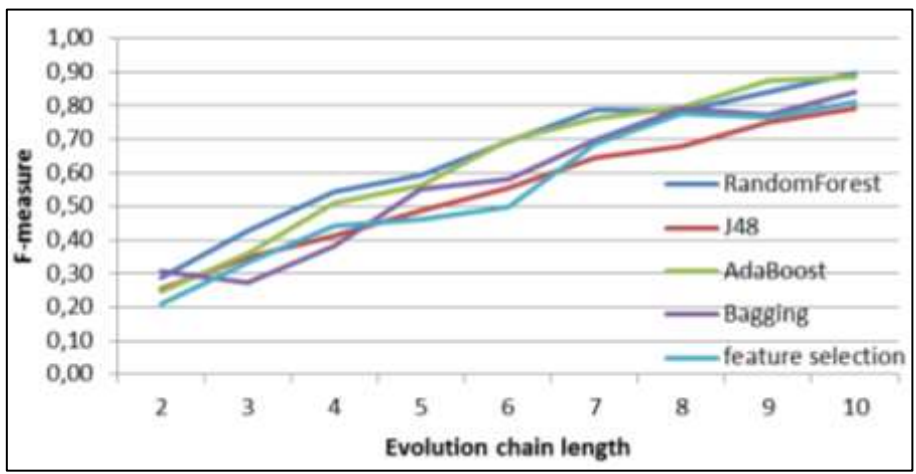

**Figure 15.** SGCI: results of event classification for *constancy* event in the Facebook dataset.

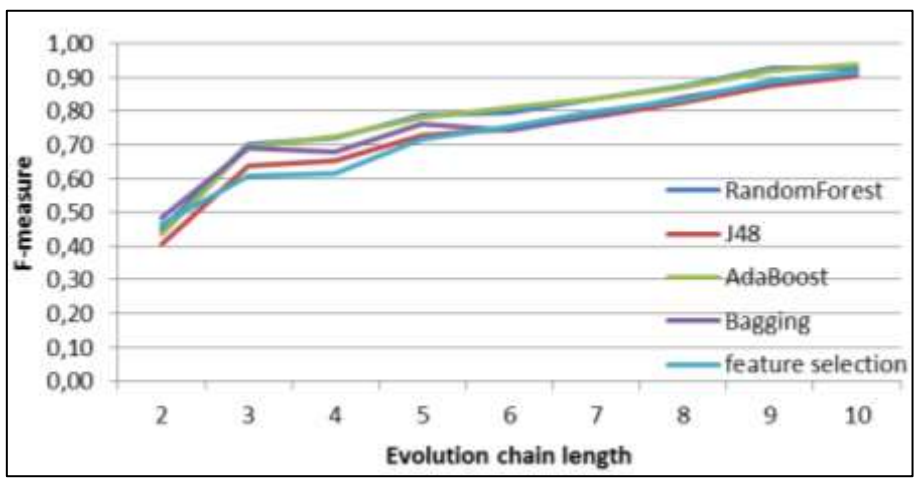

**Figure 16.** SGCI: results of event classification for *decay* event in the Facebook dataset.

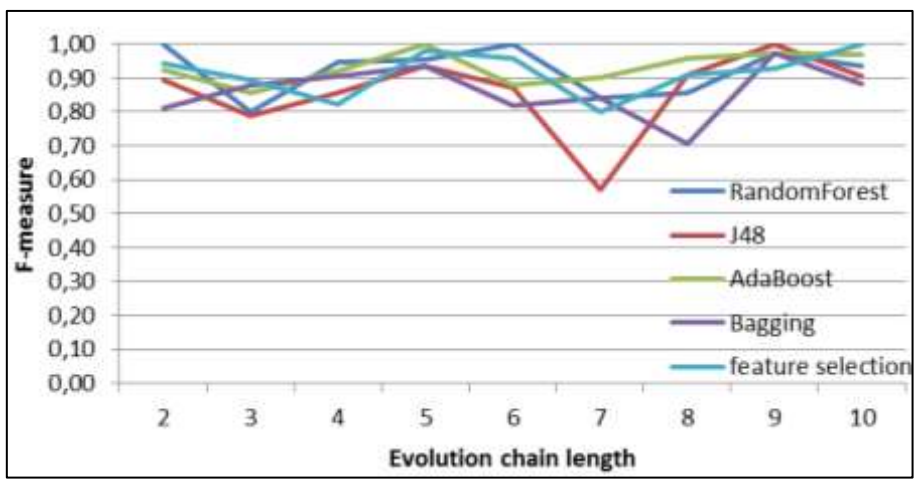

**Figure 17.** SGCI: results of event classification for *deletion* event in the Facebook dataset.

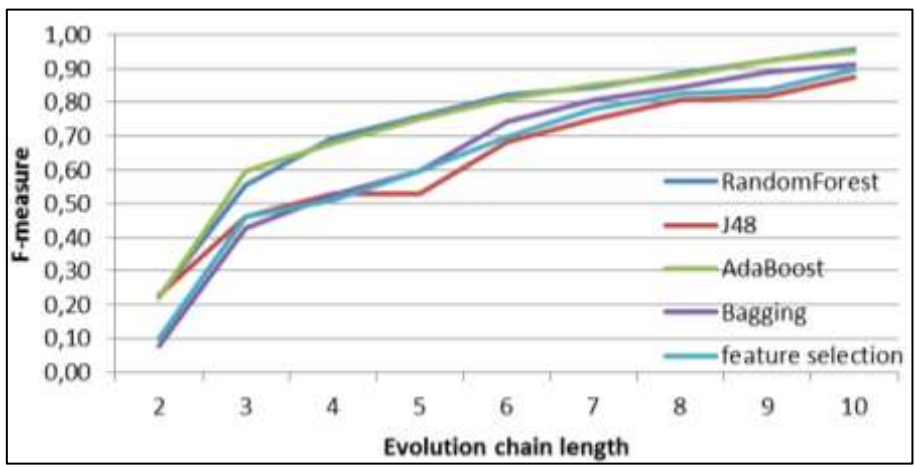

**Figure 18.** SGCI: results of event classification for *merge* event in the Facebook dataset.

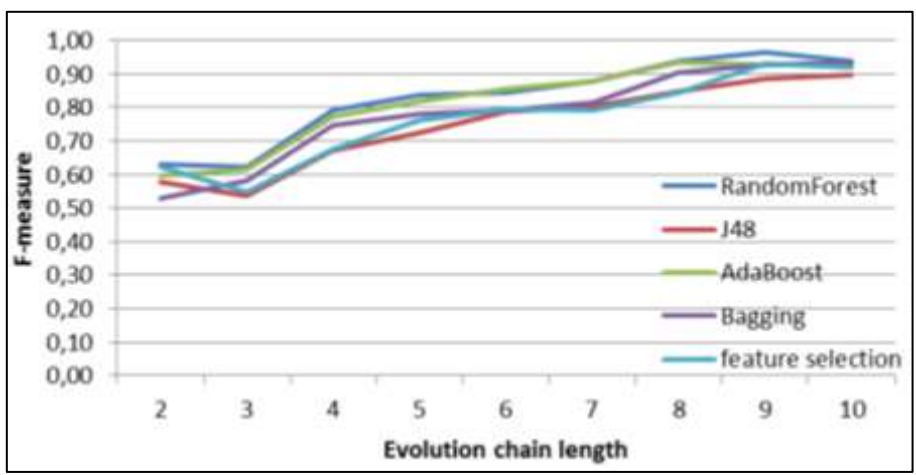

**Figure 19.** SGCI: results of event classification for *split* event in the Facebook dataset.

7.1.3. Salon24 Dataset

The best results are achieved for the Salon24 dataset. This dataset is characterised by the higher number of large groups and high dynamics caused by high activity of bloggers intensively writing post and comments.

In opposite to the described earlier datasets, the number of events is much higher for Salon24, and additionally, it rises together with the increase in the considered chain length (Figure 20, Table 5). There is a significant difference between the number of events for the shortest and longest chains (Figures 21–27). For example, for the events which have the greatest influence on a group reorganisation, e.g., *split*, *deletion*, this increase reaches almost nine and five times, respectively.

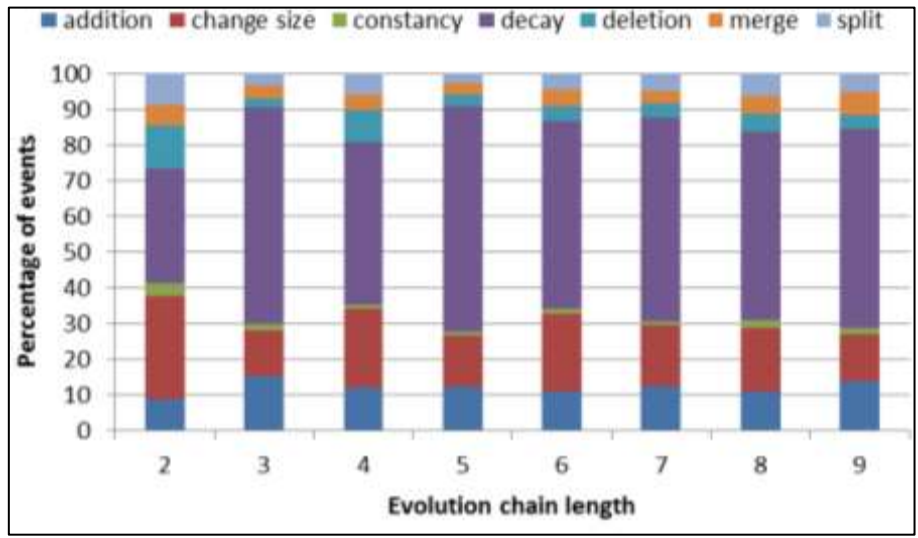

**Figure 20.** SGCI: distribution of the event types for events being predicted in the Salon24 dataset.

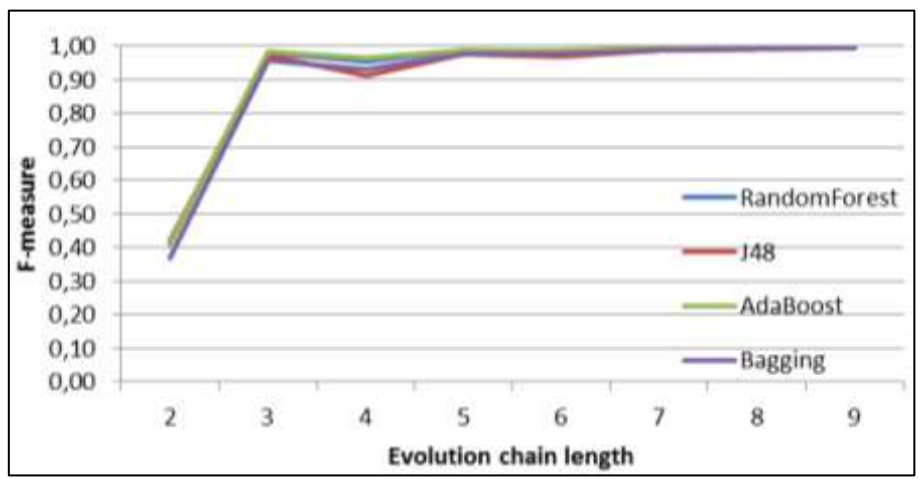

**Figure 21.** SGCI: results of event classification for *addition* event in the Salon24 dataset.

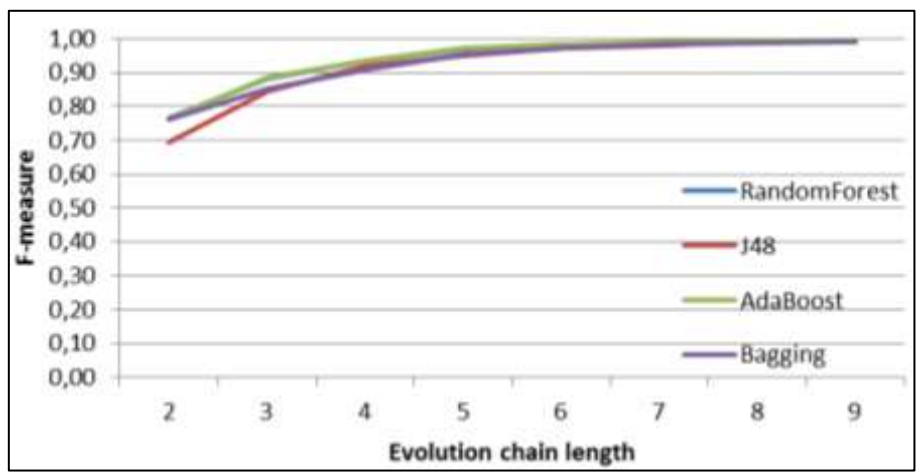

**Figure 22.** SGCI: results of event classification for *change size* event in the Salon24 dataset.

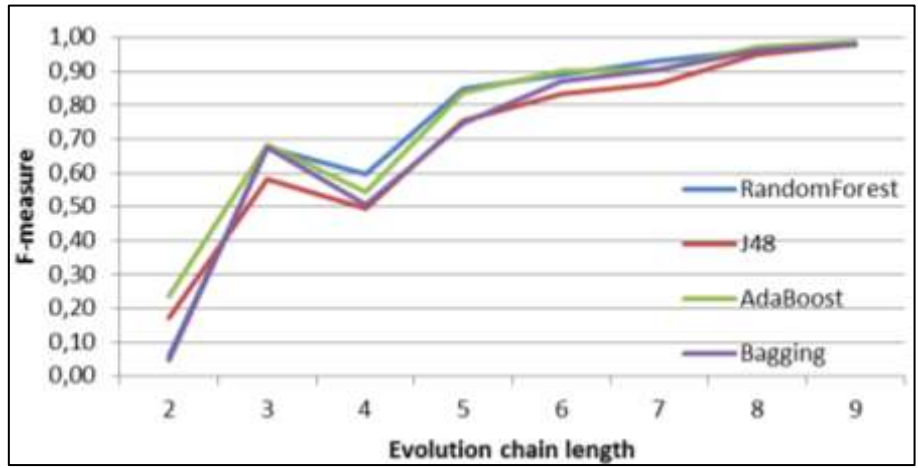

**Figure 23.** SGCI: results of event classification for *constancy* event in the Salon24 dataset.

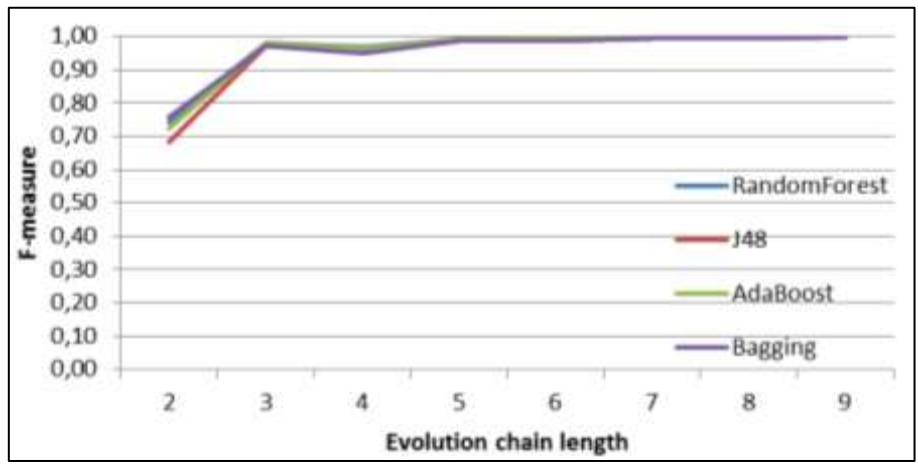

**Figure 24.** SGCI: results of event classification for *decay* event in the Salon24 dataset.

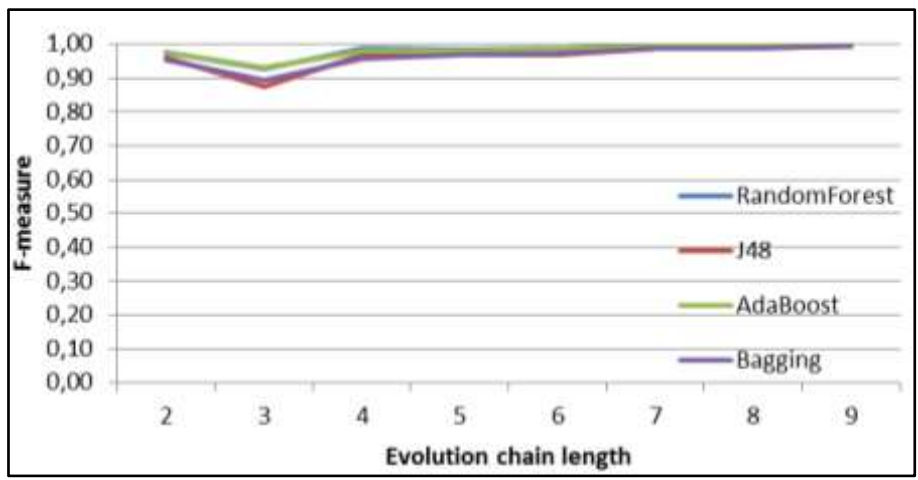

**Figure 25.** SGCI: results of event classification for *deletion* event in the Salon24 dataset.

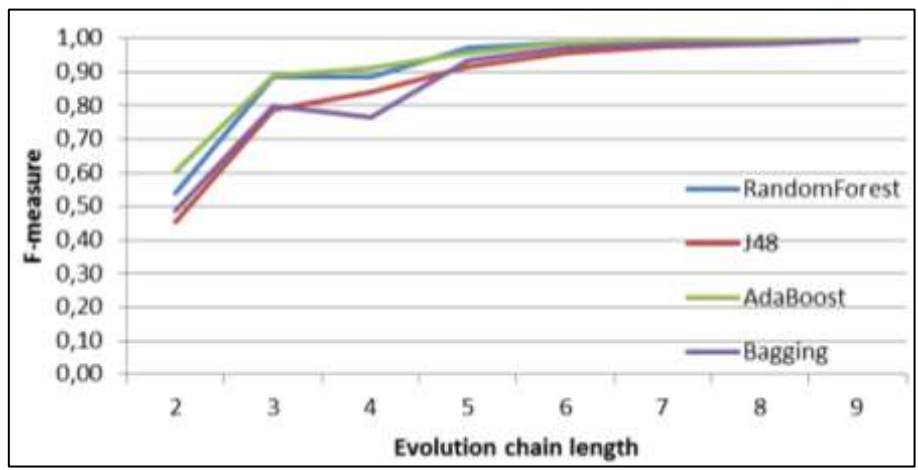

**Figure 26.** SGCI: results of event classification for *merge* event in the Salon24 dataset.

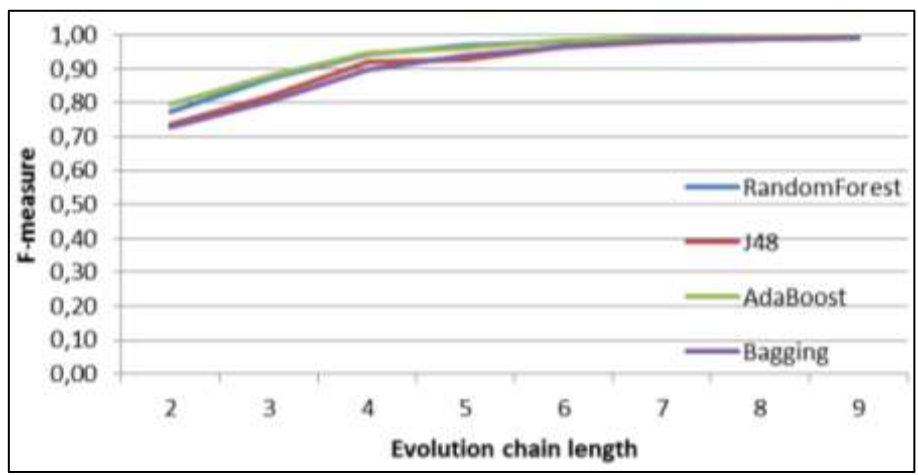

**Figure 27.** SGCI: results of event classification for *split* event in the Salon24 dataset.

### 7.1.4. Features Selection

For Facebook dataset, the Backward Feature Elimination (http://goo.gl/M4jpks) with J48 (C4.5 decision tree), was carried on. The results of this method are presented in Figures 13–19 in comparison with other approaches. One can notice that future selection gives similar results (usually better than J48 and sometimes better than Bagging, but worse than RandomForest and AdaBoost) with smaller number of features, which means that the obtained results are easier for interpretation and analysis.

Table 6 presents the number of features selected from different states taken from group history for chains with different length. The results indicate that features describing recent few states for the group have greater importance on prediction quality than the other ones. This observation is particularly evident for the longest chains.

**Table 6.** SGCI: the number of features selected for particular chain length for Facebook dataset.

| State | Chain 2 | Chain 3 | Chain 4 | Chain 5 | Chain 6 | Chain 7 | Chain 8 | Chain 9 | Chain 10 |
|-------|---------|---------|---------|---------|---------|---------|---------|---------|----------|
| n-1 | 3 | 14 | 20 | 17 | 21 | 24 | 19 | 18 | 17 |
| n-2 | 1 | 7 | 22 | 9 | 9 | 15 | 12 | 7 | 5 |
| n-3 | | 2 | 24 | 9 | 7 | 13 | 8 | 4 | 7 |
| n-4 | | | 14 | 7 | 10 | 14 | 6 | 3 | 3 |
| n-5 | | | | 3 | 9 | 14 | 7 | 6 | 5 |
| n-6 | | | | | 2 | 6 | 5 | 3 | 2 |
| n-7 | | | | | | 6 | 6 | 6 | 1 |
| n-8 | | | | | | | 1 | 3 | 2 |
| n-9 | | | | | | | | 1 | 0 |
| n-10 | | | | | | | | | 3 |

## 7.2. Predicting Group Evolution Using GED Results

The events distribution in all three examined datasets is presented in Figure 28. In the DBLP and Facebook datasets, *forming* and *dissolving* events are more frequent than others, which means that most of the groups in these networks live only for one time frame. There is a very small number of groups which last over several following time frames. On the other hand, in the Salon24 dataset *splitting* and *merging* events dominate other events; groups in this network significantly change their structure, but they last over a longer period.

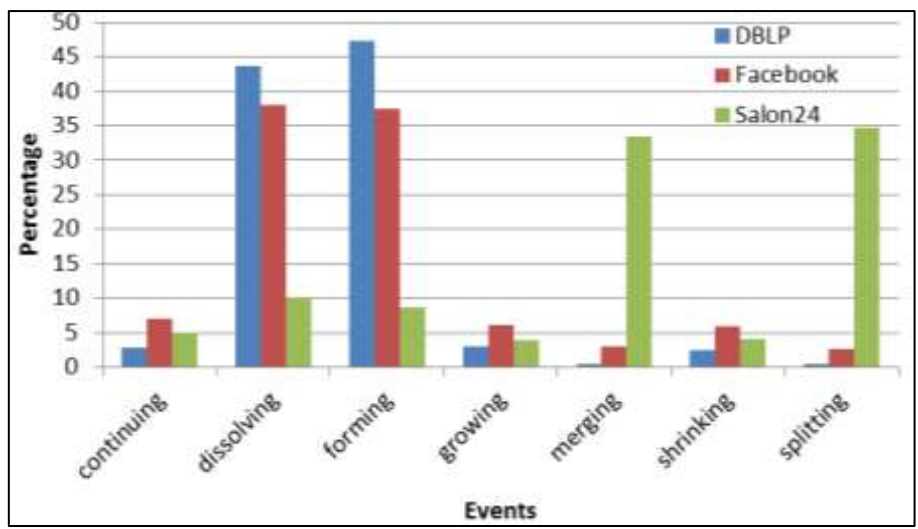

**Figure 28.** GED: distribution of the event types in all datasets.

The number of evolution chains created for particular dataset is showed in Table 7, while Tables 9–11 contain detailed statistics on the number of the particular event types being predicted. It is visible in Table 7 that the number of evolution chains for the DBLP and Facebook datasets decreases when chain becomes longer. It is a consequence of short lifetime of the groups in these datasets. The Salon24 dataset, where the *splitting* event is dominating, has the opposite tendency. Groups live longer resulting in more and more evolution chains with the increasing length. Thus, for evolution chain of length 6 only 10% and for evolution chain of length 7 only 5% of the total number of evolution chains were randomly selected (maintaining events distribution) as a input for the classifier. For the evolution chains at least as long as 8 or greater, processing of the results failed due to computational complexity, e.g., for evolution chain of length 8 there was 10 million of rows, each with 260 columns, most of them with float values.

**Table 7.** GED: the number of evolution chains for particular chain length.

| Chain Length | DBLP | Facebook | Salon24 |
|:---:|:---:|:---:|:---:|
| 2 | 20,324 | 8655 | 26,619 |
| 3 | 2480 | 3618 | 25,136 |
| 4 | 729 | 2401 | 160,059 |
| 5 | 281 | 1838 | 163,723 |
| 6 | 135 | 1434 | 107,554 * |
| 7 | 73 | 1249 | 42,284 ** |
| 8 | 45 | 1069 | - |
| 9 | 24 | 864 | - |
| 10 | 9 | 677 | - |

Note: * and ** denote that only 10% and 5% of the total number of evolution chains were selected as a input for the classifier

**Table 8.** GED: the number of evolution chains for particular event type and particular chain length in the DBLP dataset.

| Chain Length | Continuing | Dissolving | Growing | Merging | Shrinking | Splitting |
|:---:|:---:|:---:|:---:|:---:|:---:|:---:|
| 2 | 1063 | 16,875 | 1075 | 135 | 977 | 199 |
| 3 | 233 | 1557 | 285 | 69 | 229 | 107 |
| 4 | 73 | 337 | 119 | 41 | 128 | 31 |
| 5 | 26 | 113 | 51 | 15 | 56 | 20 |
| 6 | 8 | 39 | 33 | 15 | 29 | 11 |
| 7 | 4 | 16 | 18 | 6 | 21 | 8 |
| 8 | 3 | 9 | 12 | 5 | 12 | 4 |
| 9 | 1 | 9 | 6 | 3 | 4 | 1 |
| 10 | 1 | 2 | 0 | 1 | 4 | 1 |

**Table 9.** GED: the number of evolution chains for particular event type and particular chain length in the Facebook dataset.

| Chain Length | Continuing | Dissolving | Growing | Merging | Shrinking | Splitting |
|:---:|:---:|:---:|:---:|:---:|:---:|:---:|
| 2 | 915 | 4842 | 826 | 359 | 916 | 797 |
| 3 | 410 | 1193 | 512 | 257 | 642 | 604 |
| 4 | 263 | 587 | 379 | 209 | 483 | 480 |
| 5 | 191 | 388 | 300 | 160 | 399 | 400 |
| 6 | 153 | 272 | 262 | 160 | 322 | 265 |
| 7 | 129 | 205 | 218 | 124 | 259 | 314 |
| 8 | 124 | 177 | 190 | 109 | 250 | 219 |
| 9 | 89 | 176 | 149 | 121 | 166 | 163 |
| 10 | 69 | 121 | 116 | 97 | 135 | 139 |

**Table 10.** GED: the number of evolution chains for particular event type and particular chain length in the Salon24 dataset.

| Chain Length | Continuing | Dissolving | Growing | Merging | Shrinking | Splitting |
|---|---|---|---|---|---|---|
| 2 | 115 | 341 | 114 | 957 | 142 | 24,950 |
| 3 | 214 | 1179 | 230 | 10,517 | 249 | 12,747 |
| 4 | 112 | 727 | 123 | 5632 | 183 | 153,282 |
| 5 | 1090 | 8724 | 1019 | 66,511 | 1542 | 84,837 |
| 6 * | 60 | 593 | 62 | 3808 | 111 | 102,920 |
| 7 ** | 591 | 878 | 573 | 17,958 | 317 | 21,967 |

Note: * and ** denote that only 10% and 5% of the total number of evolution chains were selected as a input for the classifier

The F-measure value for each dataset, each event type and all classifiers is presented in Figures 30–35,37–42,44–49. As in case of *SGCI* method for all datasets classifiers RandomForest, J48, AdaBoost and Bagging were used. Additionally for Facebook dataset (Figures 37–42) Feature Selection mechanism was used.

### 7.2.1. DBLP Dataset

For the first dataset, DBLP, GED parameters were lowered to 30% in order to not omitting small groups, which are majority of the DBLP dataset (Figure 4). For this dataset evolution chains of length from 2 to 10 were selected. While increasing the length of the chain the total number of evolution chains was decreasing (Table 8).

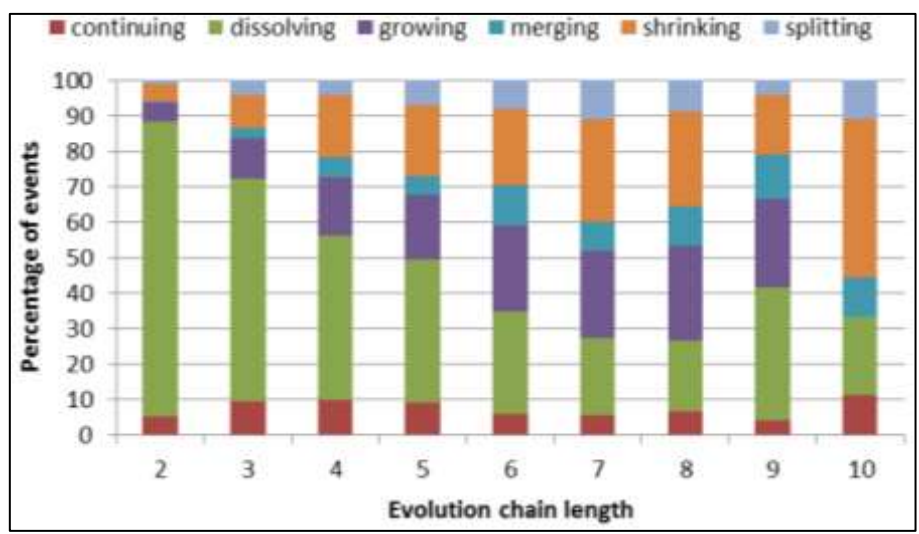

**Figure 29.** GED: distribution of the event types for events being predicted in the DBLP dataset.

For evolution chain of length 5, there were less than 300 evolution chains which resulted in failure of the Bagging classifier calculations for merging event (Figure 33). Results for evolution chains of length 5 and longer should not be considered and are presented only demonstratively.

The F-measure comparison for all event types and all classifiers is presented in Figures 30–35. Each figure depicts F-measure value in relation to different evolution chain lengths for one specific event type and all classifiers. It can be observed that until evolution chain of length 4, the F-measure value has tendency to grow. After that point the classifiers cannot handle too small amount of training data and provide poor results.

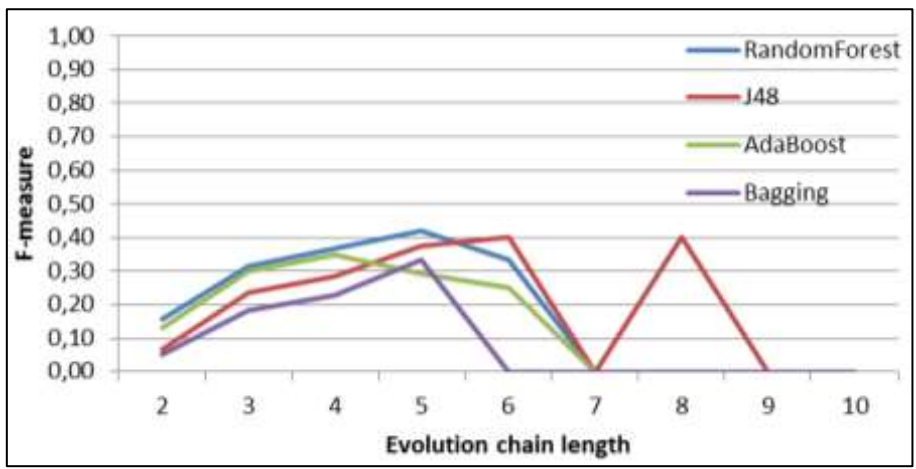

**Figure 30.** GED: results of event prediction for *continuing* event in the DBLP dataset.

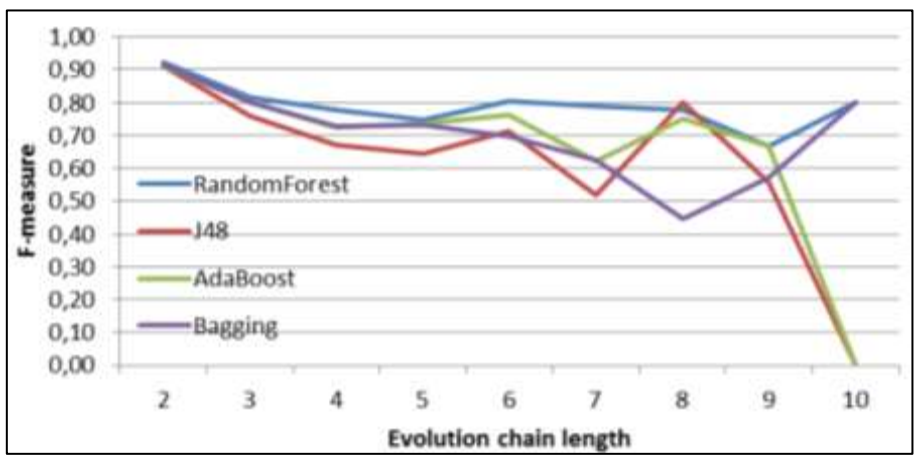

**Figure 31.** GED: results of event classification for *dissolving* event in the DBLP dataset.

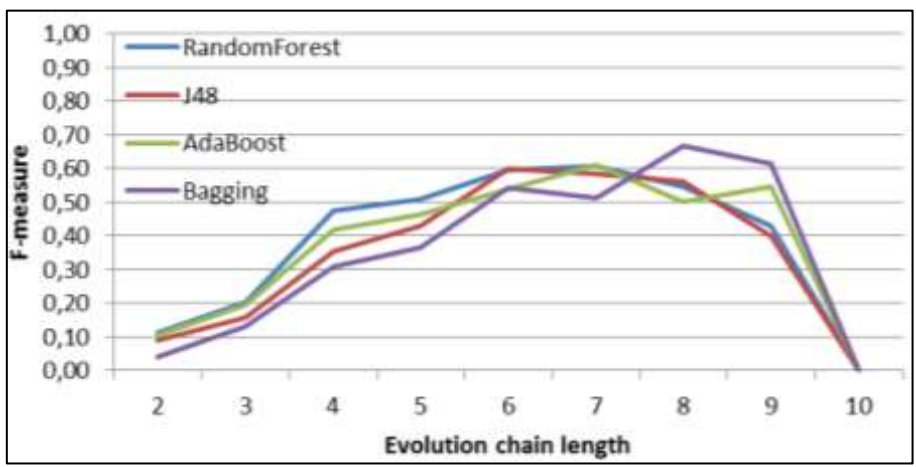

**Figure 32.** GED: results of event classification for *growing* event in the DBLP dataset.

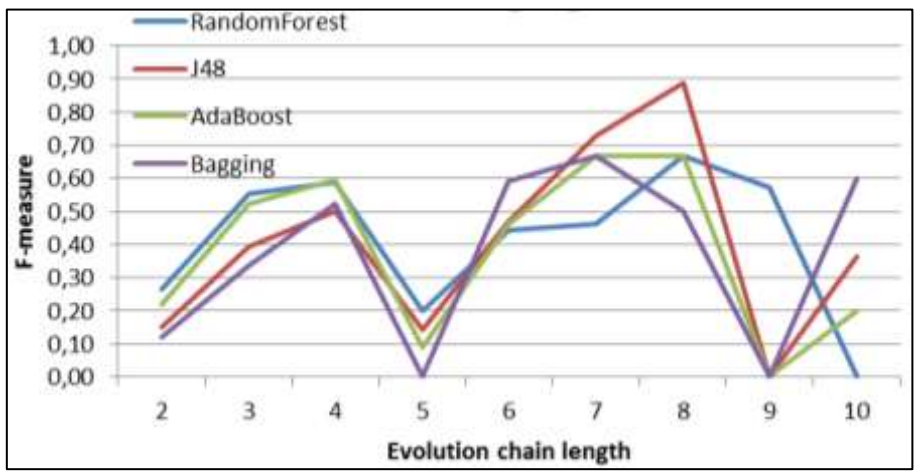

**Figure 33.** GED: results of event classification for *merging* event in the DBLP dataset.

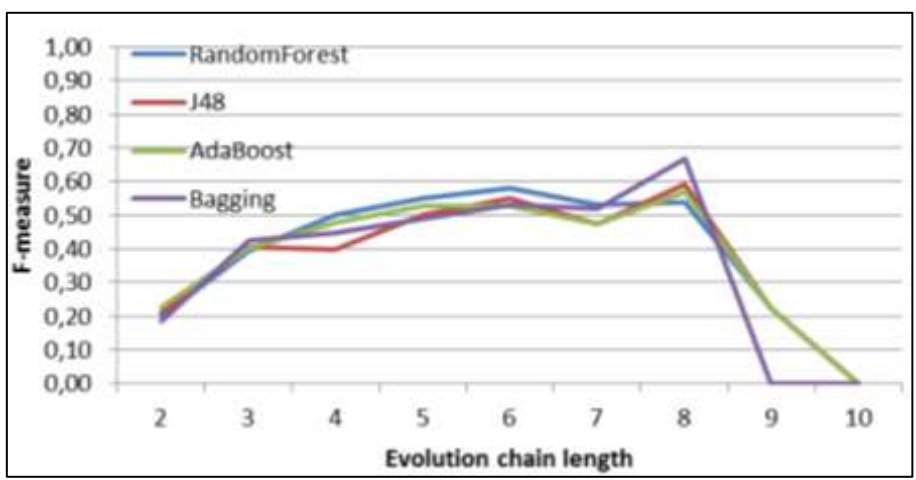

**Figure 34.** GED: results of event classification for *shrinking* event in the DBLP dataset.

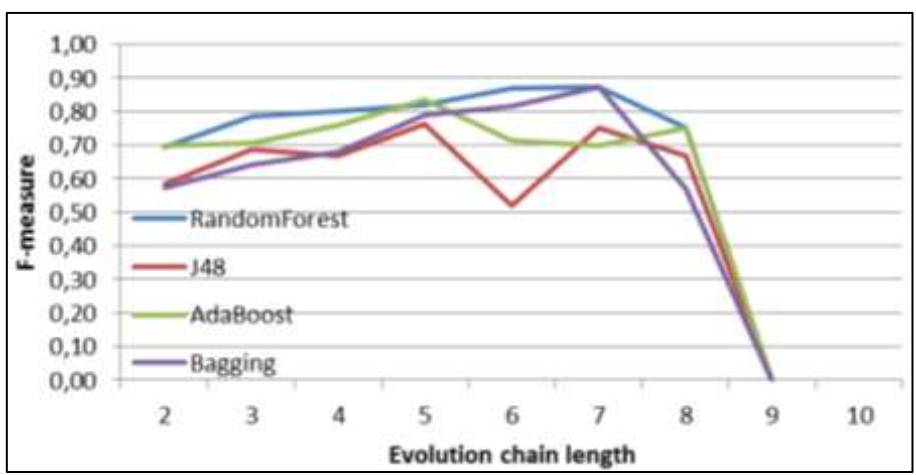

**Figure 35.** GED: results of event classification for *splitting* event in the DBLP dataset.

Figure 31 presents decreasing value of F-measure while the length of evolution chain increases. It is connected to events distribution (Figure 29). For evolution chain of length 2 dissolving event occurs much more often than other events (83% of all events), however with longer evolution chains domination of dissolving event decreases (63% for chains of length 3 and 46% for chains of length 4).

The Bagging classifier seems to be more vulnerable to the insufficient number of training data than other classifiers.

### 7.2.2. Facebook Dataset

For the second dataset—Facebook, the GED parameters were also lowered to 30% in order to cope with unstable network. Again evolution chains of length from 2 to 10 were selected. Similarly to DBLP dataset, while increasing the length of the chain the total number of evolution chains was decreasing (Table 9).

Apart from running four classifiers, Backward Feature Elimination block from KNIME was used. The basic classifier used to select features was J48 decision tree. The comparison between classifiers with all features and J48 with feature selection is presented in Figures 37–42. Each figure shows F-measure value across different evolution chain lengths for one specific event type. Aside from *dissolving* event relationship between longer evolution chains and higher value of F-measure can be seen as in case of DBLP dataset.

The F-measure for *dissolving* event is initially at level of 0.8 but then drops below 0.7 (Figure 38). It is again related to the events distribution, where for evolution chain of length 2 *dissolving* has great advantage over other events but while evolution chains are getting longer the events distribution becomes flat (Figure 36).

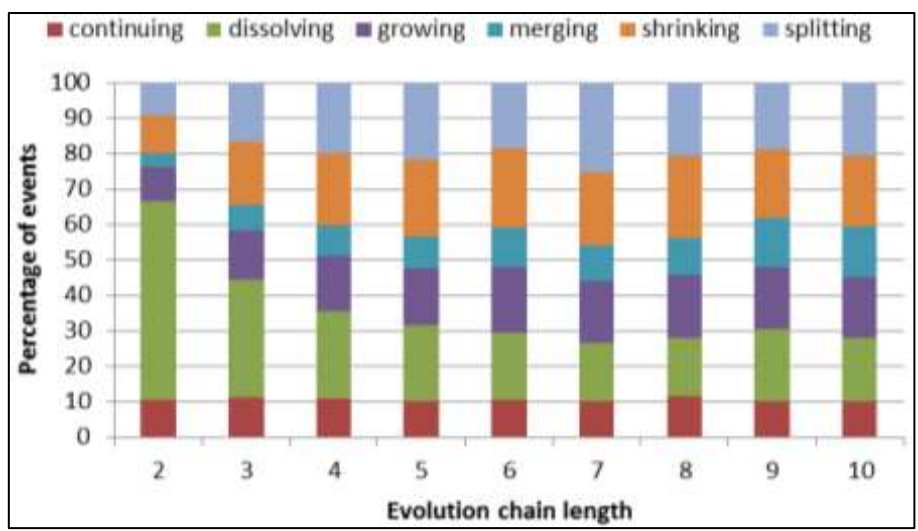

**Figure 36.** GED: distribution of the event types for events being predicted in the Facebook dataset.

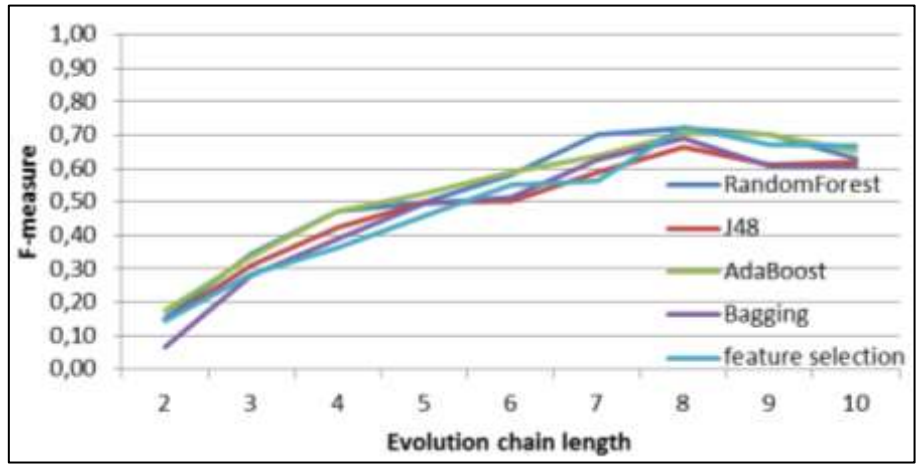

**Figure 37.** GED: results of event classification for *continuing* event in the Facebook dataset.

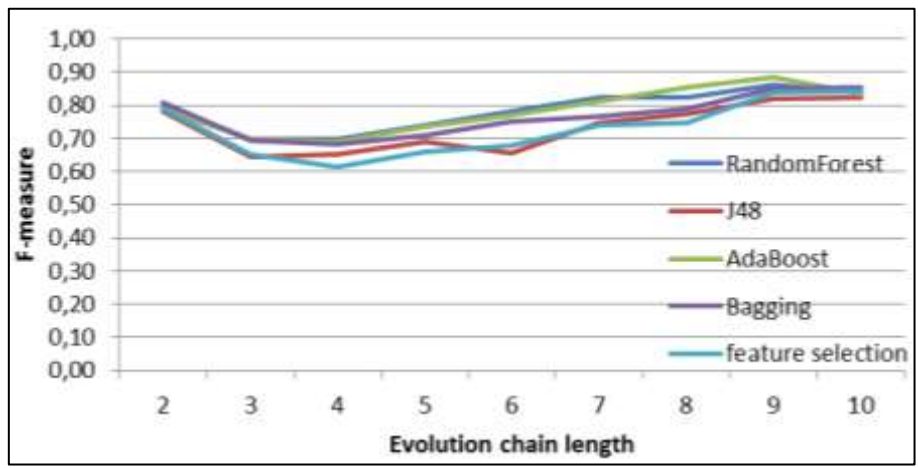

**Figure 38.** GED: results of event classification for *dissolving* event in the Facebook dataset.

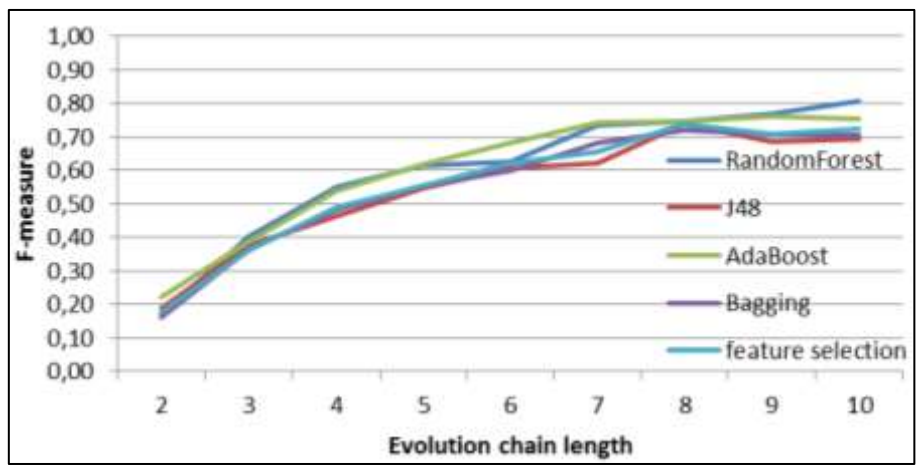

**Figure 39.** GED: results of event classification for *growing* event in the Facebook dataset.

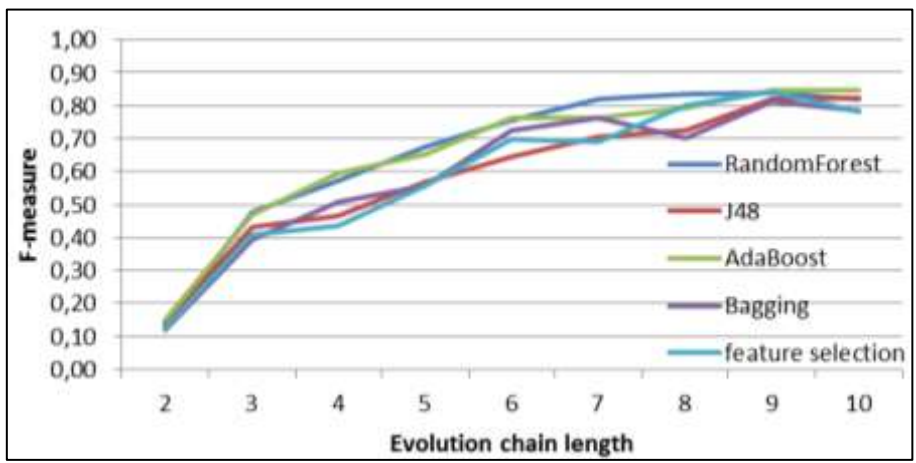

**Figure 40.** GED: results of event classification for *merging* event in the Facebook dataset.

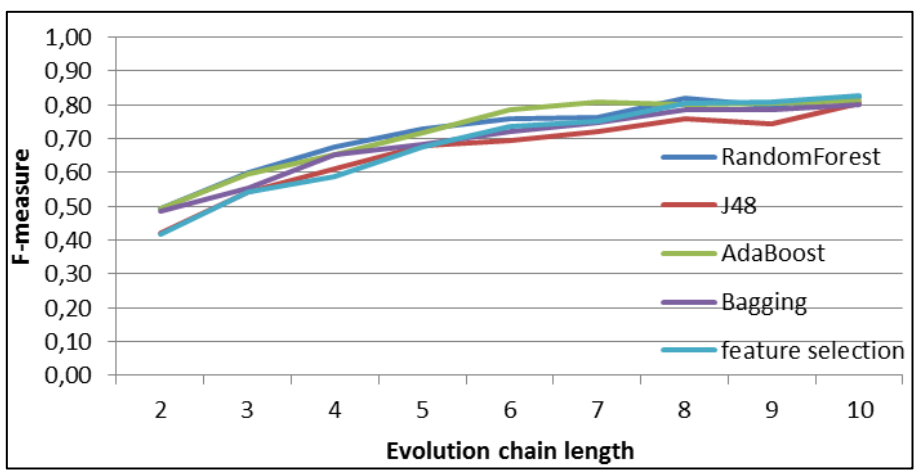

**Figure 41.** GED: results of event classification for *shrinking* event in the Facebook dataset.

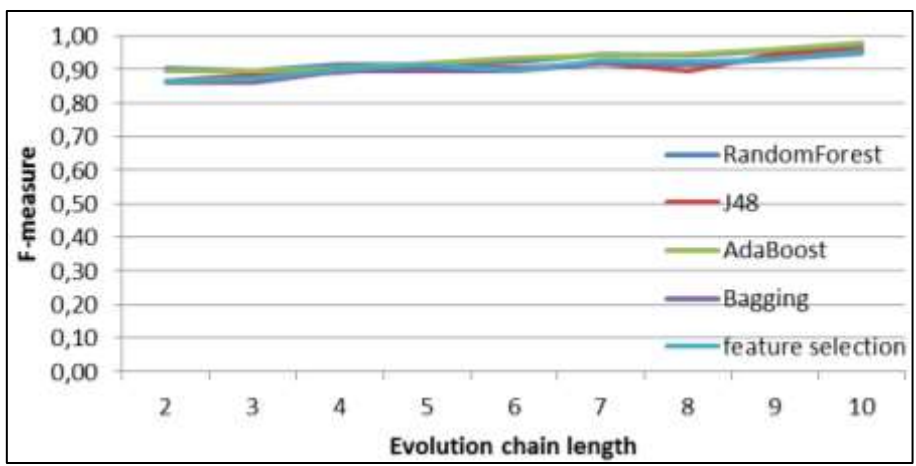

**Figure 42.** GED: results of event classification for *splitting* event in the Facebook dataset.

For evolution chains of length 8 and longer the value of F-measure is not gaining as significantly as for shorter chains. This might be due to the less number of evolution chains in training dataset for classifiers. It is impossible to state whether J48 classifier or J48 preceded by feature selection mechanism gives better results. In general results for all classifiers are very similar. Broader analysis on feature selection is described in further Section 8.2.

### 7.2.3. Salon24 Dataset

For the last dataset, Salon24, the GED parameters equal 70 where utilized, which means high similarity between groups in consecutive time frames within the same evolution chain.

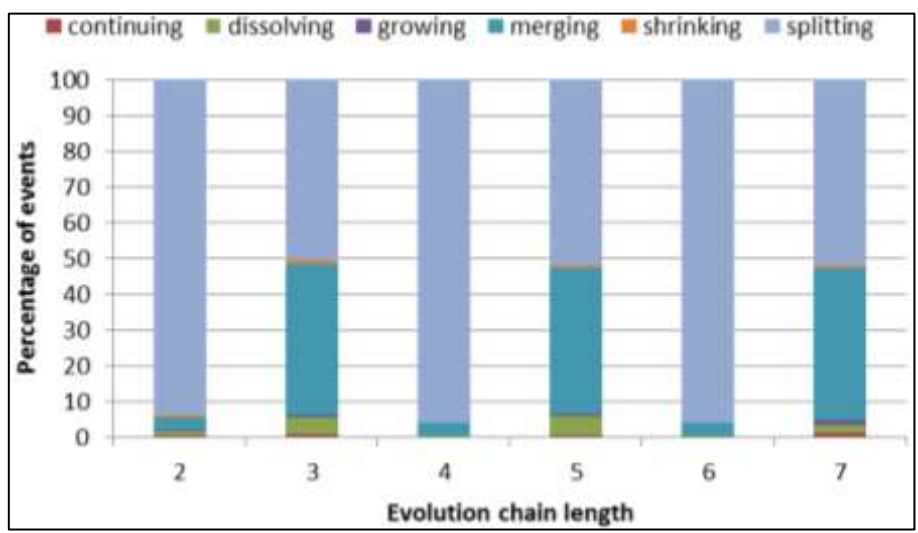

**Figure 43.** GED: distribution of the event types for events being predicted in the Salon24 dataset.

For Salon24 dataset evolution chains of length from 2 to 7 were selected. In opposite to previous datasets, while increasing the length of the chain the total number of evolution chains was rapidly growing (Table 10).

Figures 44–49 show the F-measure value behaviour while increasing the length of evolution chain for one specific event type and all classifiers. The figures clearly show that the value of F-measure grows with the length of the evolution chain. The exception is *splitting* event for which the value of F-measure reaches almost 1 already for evolution chain of length 2. For evolution chain of length 7 the value of F-measures is close to 1 for all types of events and all classifiers.

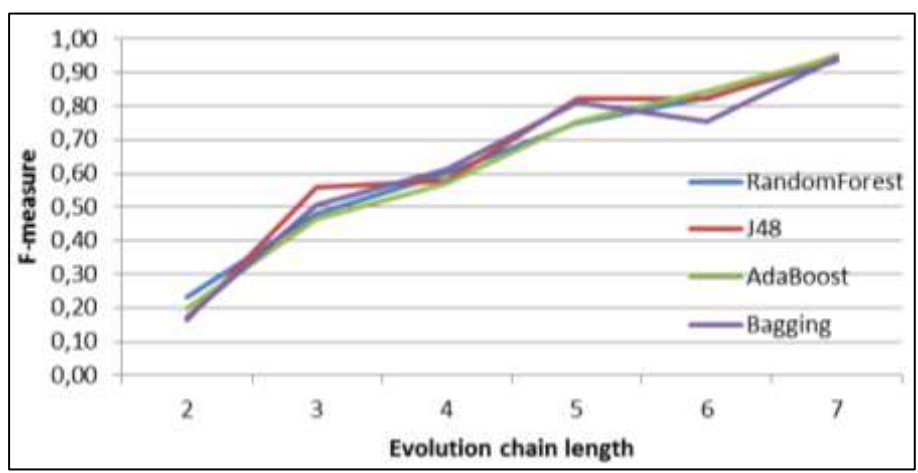

**Figure 44.** GED: results of event classification for *continuing* event in the Salon24 dataset.

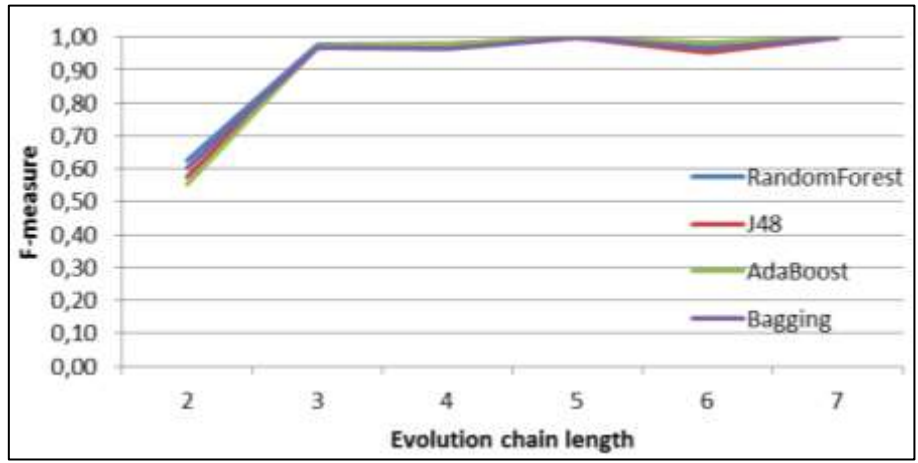

**Figure 45.** GED: results of event classification for *dissolving* event in the Salon24 dataset.

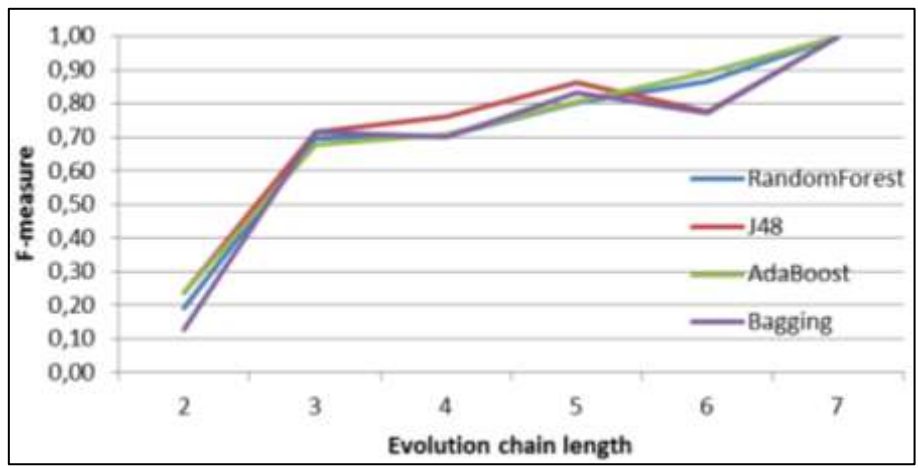

**Figure 46.** GED: results of event classification for *growing* event in the Salon24 dataset.

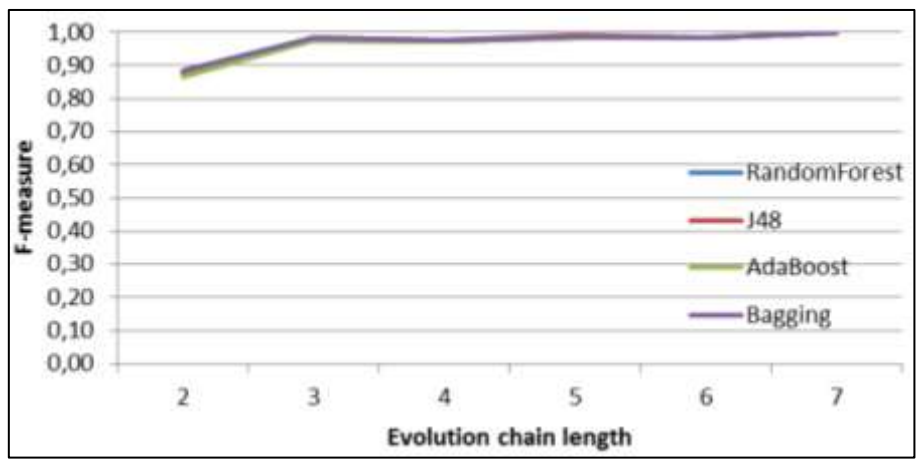

**Figure 47.** GED: results of event classification for *merging* event in the Salon24 dataset.

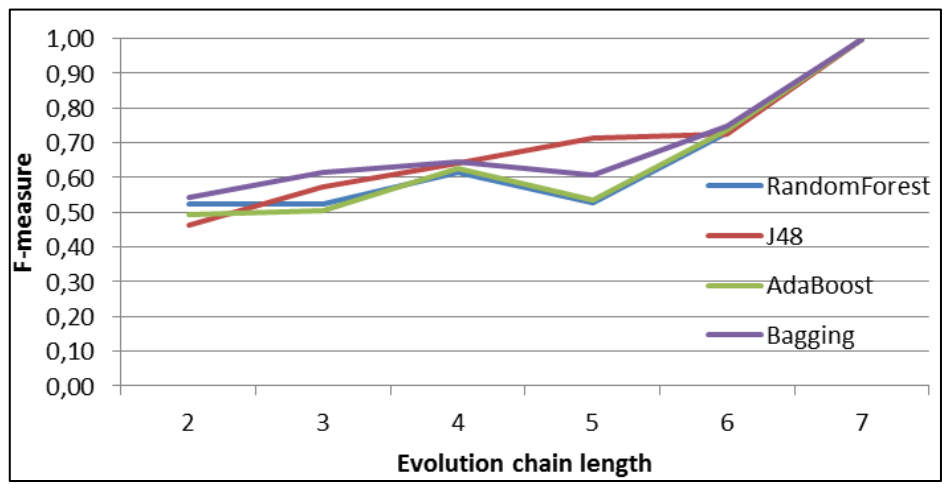

**Figure 48.** GED: results of event classification for *shrinking* event in the Salon24 dataset.

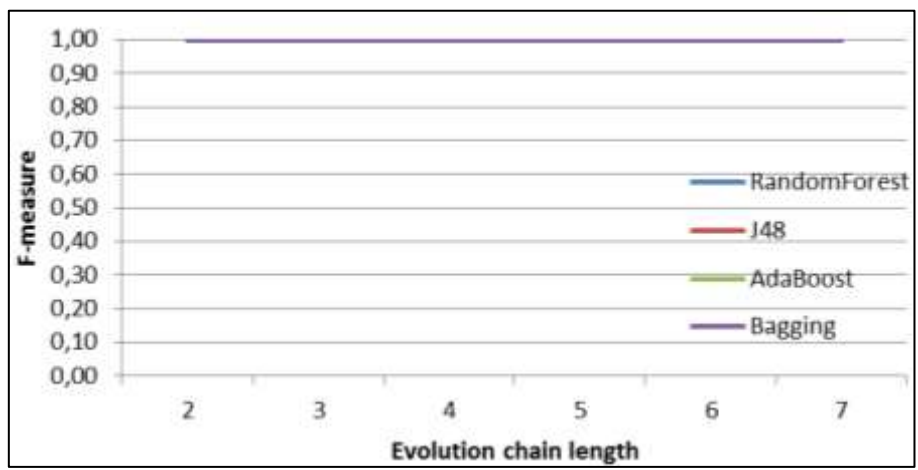

**Figure 49.** GED: results of event classification for *splitting* event in the Salon24 dataset.

As we can see each classifier achieves better results for *splitting*, *merging* and *dissolving* events and worse for *continuing*, *shrinking* and *growing*. This happens because of uneven distribution of different event types (Figure 43). The number of splitting *events* is much higher than the number of the other events. We think this is because the time frame size is too short for the most communities and they continuously *splits* and *merge* as service users migrates from one topic to another. Authors of the GED method suggests in [23] that increasing the size of the time frame increases the possibility for the emergence of persistent groups and this will be our next step in future work.

7.2.4. Features Selection

Table 11 provides the number of features selected as an input for classifier. To determine which features should be selected Backward Feature Elimination mechanism implemented in KNIME was used. As the results shows for longer evolution chains less features were selected. Broader analysis of the results are in Section 8.2.

**Table 11.** GED: the number of features selected for particular chain length for the Facebook dataset.

| State | Chain 2 | Chain 3 | Chain 4 | Chain 5 | Chain 6 | Chain 7 | Chain 8 | Chain 9 | Chain 10 |
|-------|---------|---------|---------|---------|---------|---------|---------|---------|----------|
| n-1   | 12      | 21      | 23      | 10      | 16      | 26      | 12      | 14      | 14       |
| n-2   | 6       | 25      | 22      | 8       | 9       | 15      | 5       | 9       | 7        |
| n-3   |         | 8       | 24      | 10      | 11      | 8       | 7       | 4       | 3        |
| n-4   |         |         | 5       | 5       | 3       | 7       | 6       | 5       | 2        |
| n-5   |         |         |         | 0       | 7       | 7       | 6       | 1       | 0        |
| n-6   |         |         |         |         | 0       | 3       | 6       | 2       | 0        |
| n-7   |         |         |         |         |         | 0       | 4       | 1       | 0        |
| n-8   |         |         |         |         |         |         | 1       | 1       | 1        |
| n-9   |         |         |         |         |         |         |         | 0       | 0        |
| n-10  |         |         |         |         |         |         |         |         | 0        |

## 8. Discussion

### 8.1. Prediction

The quality of prediction was evaluated using three datasets with different profiles (see Section 6). These differences can be especially perceived in the context of group characteristics and their dynamics. By dynamics of groups we mean both, the change in group profile and change of events.

For both methods (SGCI and GED) the best results are achieved using RandomForest and AdaBoost classifiers while J48 and Bagging classifiers provided slightly worse results. However, the differences in F-measure value between all classifiers are not large, which is visible in all Figures from the previous Section 7. Sometimes, it is even hard to point out which classifier is better (see e.g., Figure 42), yet in general, the ensemble classifiers outperform the others.

The clear observation from the charts presented is that the increasing length of the evolution chain improves prediction accuracy. It is hard to unequivocally state why this happens. The most intuitive and the first explanation tells that the longer evolution chains, the broader and more comprehensive history with more information about the group. In consequence the classifier has more potentially valuable features to predict the future event. However, the results obtained while applying the features selection mechanism also tend to have higher value of F-measure for longer evolution chains (Figures 13–19 and 37–42), even though the number of features selected to predict the future event is not increasing with the length of the evolution chain (Figures 50 and 51). Moreover, the features from the last three states (time frames) are selected most frequently (Figure 52), for details see Section 8.2.

We can still try to consider why the results are better for longer evolution chain. Maybe, it is because the general number of evolution chains is greater for longer chains (Salon24, Tables 3 and 8)? Regrettably, the other datasets do not match this explanation. Perhaps the event distribution in the training dataset becomes more balanced for longer evolution chains and this provides more accurate prediction? Unfortunately not, since this is valid only for the Facebook dataset and the GED method (Figure 36). This also can hardly result from the profile of the dataset since all three datasets are different and for all three the property "the longer chain, the better results" is visible.

Despite several hypotheses, there is no clear and fully reliable explanation why the increasing length of the evolution chain improves the prediction accuracy. This issue requires further studies that will be carried out in the future.

Another interesting question is why the results for the Salon24 dataset are better than results for DBLP and Facebook? It is particularly visible for the GED method: compare equivalent Figures 44–49 with Figures 30–35 and 37–42. It may be related to the smaller number of evolution chains for datasets DBLP and Facebook (Table 7). However, the number of chains for the DBLP and Salon24 datasets are almost the same for their length of two (Table 7) and the results for Salon24 are still better. Furthermore, the event distributions are very similar–one event far outweighs the other events (Figures 29 and 43). Anyway, the type of the event may be also important. Event *dissolving* dominates other events in DBLP whereas *split* is most frequent for the Salon24 dataset.

This leads to a general conclusion that the profile of the dataset really matters. The DBLP network consists of small groups which are alive for a small number of time frames. In the Facebook network groups are even smaller but survive a little bit longer. Finally, the Salon24 network is compounded of larger groups, which change their structure very often but they last over many time frames. It is easier to predict the future for long lasting groups.

For the SGCI method, one can notice that an increase in chain length usually gives the better quality of prediction as long as the accuracy is not close to its maximum possible value. The possible explanation for this fact may be derived from the way the evolution chains are being built, *i.e.*, for every possible path in group history starting from a given past group state to its current state, it is likely to pass through many different chains. Moreover, it may happen that some of these chains (for the same group) are in the training part of the dataset and some of them are used for testing. If classifiers use only features from the common part of these chains, they may produce too optimistic results. Typically, for many cases, especially for the SGCI method, there exists a certain chain length, for which the prediction quality (F-measure) is close to its maximum value. It means that the prediction quality cannot be significantly improved by increasing the evolution chain length. For example, this border chain length is four time frames for DBLP (Figures 7–11), and the same for *addition* event for Facebook (Figure 13). For Salon24, this point is at the level of three for events *addition* and *decay* (Figures 21 and 24), four in the case of *deletion* and *split* (Figures 25 and 27), and five for *change size*, *constancy* and *merge* (Figures 22,23,26).

For the GED method, the quality of results can grow with the chain length to a given point and may drop for very long chains. However, it is valid only for the DBLP dataset, see the lower quality for lengths of 5–10 in Figures 30–35. It comes evidently from the very low quantity of long length chains (<30) for the DBLP dataset, see Table 7. For the other datasets, the increase in quality for longer chains is very stable, Figures 37–42 and 44–49 because the number of chains is much more balanced. Additionally, the usability of features extracted from older time frames is much lower than for younger ones, see the next section. This is the real reason why the increase in the evolution chain length for longer values not significantly improves the prediction quality.

Takaffoli *et al.* in [18] tested their method of prediction of communities evolution on the Enron and DBLP datasets. In this paper we conducted experiments on Facebook, DBLP and Salon24, so we can only compare our results on DBLP dataset, but such comparison is limited because the method of Takaffoli differs in many ways from GED and SGCI. Moreover, we used in experiments 20 slots of length one year, but authors in [18] carried out experiments on 10 slots of length equal to one year. The values of F-measure for DBLP achieved in Takaffoli method are similar to those best achieved using GED method (but with different values of chain lengths)–split event on the level about 0.8, merge event–0.6, survive (no strict equivalent in GED method)–0.6. For SGCI those results are better-for chains of length 4 those values are about 0.9 and are rising along with growing length of chains. In Takaffoli method [18] they also consider RSurvive event which referred to Survive events assigned only to groups which existed longer than one time slot. Interesting observation is that the F-measure values for RSurvive are higher than Survive. In SGCI there is similar concept but on the level of whole method and in experiments we assigned events to groups which existed at least in three time slots., but impact of this requirement for groups on results of prediction is limited only to chains with length equal two (because procedure of creation chains with length e.g., five considers only groups lasting at least in five subsequent timeframes).

### 8.2. Features Selection

The Backward Feature Elimination algorithm with J48 tree for all datasets has been applied for both methods (SGCI–Table 6 and GED–Table 11). Results for all used datasets differ, but some measures are repeatedly more often used than others and the tendency of choosing features in different chains is similar. Therefore, in this paper we present in detail only results for Facebook dataset. The Backward Feature Elimination algorithm extracts the best set of features for each chain length. The number of used features increases until the chain with the length equal to seven (with the exception of chain of length two) and then decreases gradually.

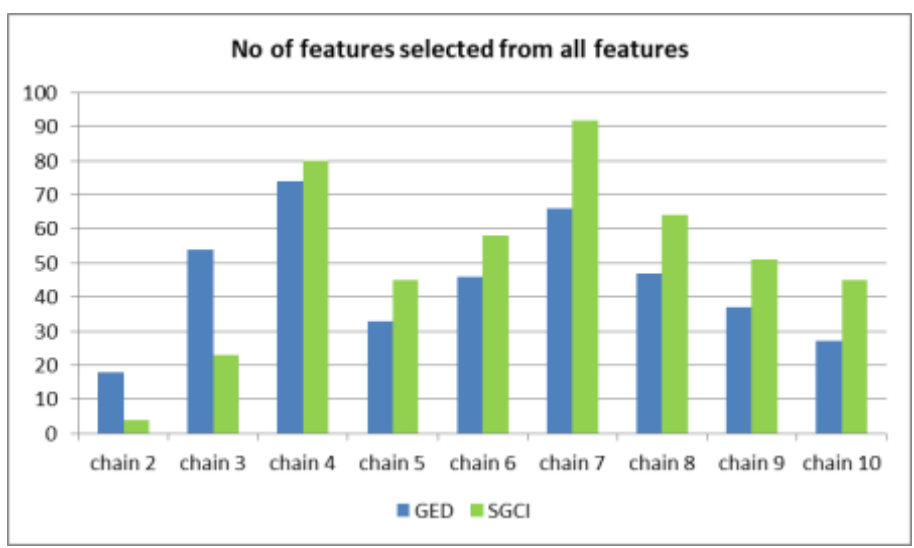

**Figure 50.** The number of features selected from all features for the particular chain length for the Facebook dataset.

Since the number of features changes with the chain length, the percentage of total features selected by the algorithm was also analysed (Figure 51). It shows the same trend even more clearly. The intuition behind this is that the knowledge contained with the events chain fluctuation is enough for the classifier to learn and it does not need so many additional information to predict the future change. In general, SGCI takes into account more features for prediction than GED. It may be connected with more events to predict in SGCI method. Furthermore, the difference in features usage is the highest for the chain equal to two and possible explanation for this fact can be the concept of stable groups in SGCI which only affects results on chain of length 2 (the number of stable groups is smaller than the number of all groups and, therefore, smaller number of features may be sufficient).

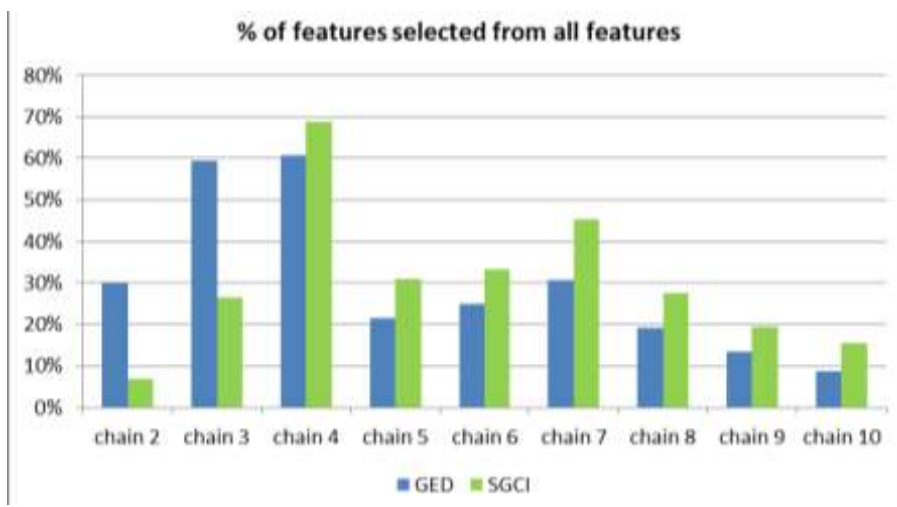

**Figure 51.** The percentage of features selected from all features available for the particular chain length for the Facebook dataset.

Moreover, we have been able to notice another interesting behaviour. When, we checked how "old" are the features selected by the feature selection algorithm, we found out that most of them are from the last three time frames (see Figure 52).

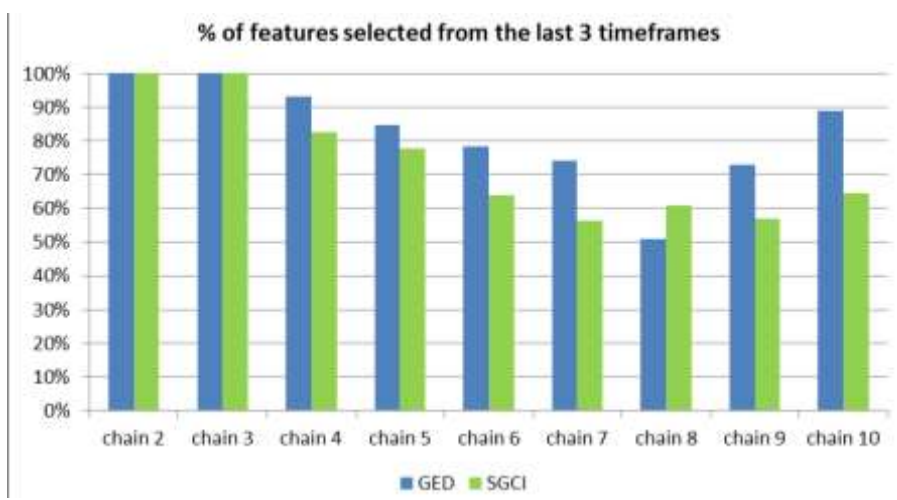

**Figure 52.** The percentage of features selected from the last 3 time frames for the particular chain length for the Facebook dataset.

When we predict event $T_n \rightarrow T_{n+1}$, most of features is from the group profiles extracted in states $T_n$, $T_{n-1}$ and $T_{n-2}$, see Figure 3 for the sequence of events for a single group. For example, when the evolution chain length was 10 and upcoming change was predicted as many as 89% of features in case of GED and 64% in case of SGCI have been from the tenth, ninth and eighth group profile.

Next, the effectiveness of all features was analysed to check if some of them are better for predicting community evolution (see Figure 53). For GED the most frequently used feature was *size*, for *SGCI size* was also chosen similar times, but there were some features more frequently used such as *min_degree_in*, *min_betweenness*, *{avg,max,min}_eigenvector* or *{avg,max}_closeness*. In general, SGCI selected more minimum and sum values of centrality metrics in groups, but GED–more average values of such metrics in groups. Another observation is that eigenvector centrality and closeness is much more often used in SGCI than in GED.

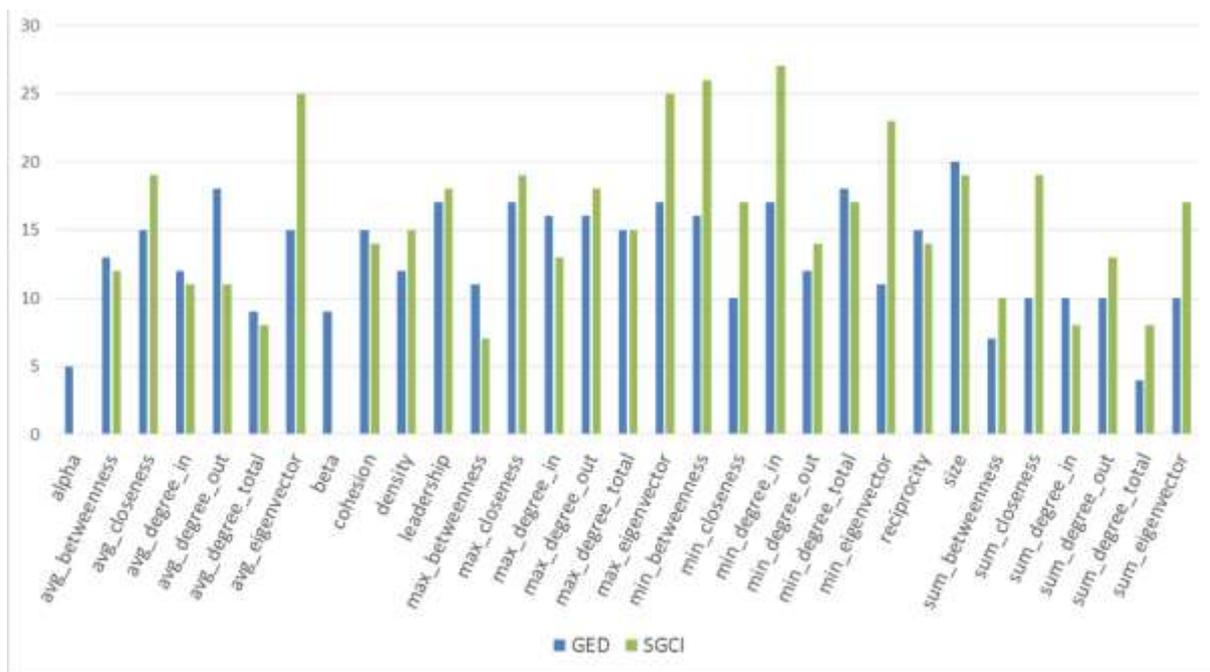

**Figure 53.** The comparison of features usage in GED and SGCI after feature selection for the Facebook dataset.

We decided also to compare errors of prediction of GED and SGCI events after feature selection (see Figure 54). The results for both methods are similar. At the beginning, errors of GED are smaller than SGCI ones, but starting from chain 4 the situation is different—SGCI has smaller errors than GED. Combining our results with previous observations regarding number of features selected by both methods, we can notice that, generally, SGCI uses more features than GED, but GED has bigger errors than SGCI.

If we combine two main conclusions from this section, *i.e.*, the longer chain the better results and that three last (youngest) windows are the most influential on the classifier results, we can conclude that we need the long history of changes but the profile for only few last group states. This could be useful if one has limited computational capabilities and cannot calculate group profiles for all groups.

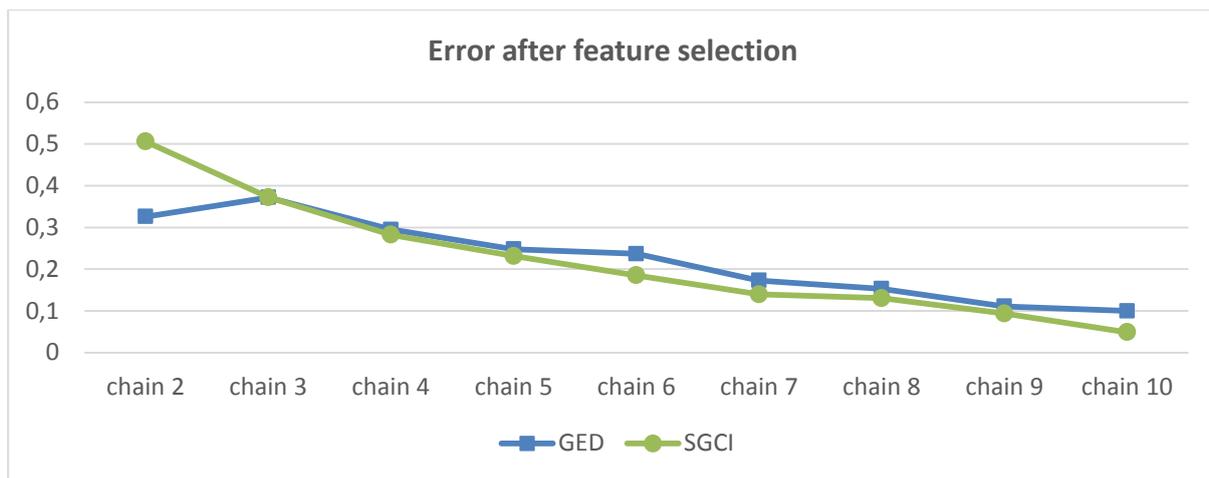

**Figure 54.** Error in prediction of *GED* and *SGCI* events after feature selection for the Facebook dataset.

## 9. Conclusions and Future Work

In this paper two methods to predict the nearest future of the social group are presented and analysed. The first method utilizes the *Stable Group Changes Identification* algorithm (SGCI) while another one makes use of the *Group Evolution Discovery* algorithm (GED). Both methods differ in approach to community identification (fugitive or stable), event definition and kind of information delivered to learn and test classifier: either only recent and previous group attributes (SGCI) or also attributes describing previous group changes (the *inclusion* measure in GED).

The experimental research was conducted on three real datasets with different characteristic: DBLP, Facebook, Salon24–Polish blogosphere, see Sections 7.1 and 7.2. Four classifiers and an additional Backward Feature Elimination mechanism were used in the process of predicting future event for the group. Evolution chains of length from two to ten time frames were evaluated as a group history. Each evolution chain contained information about previous group profile at that time, *i.e.*, features such as group size, group density, group leadership, average degree, *etc.*, as well as transitive changes between following time frames (only GED), see Section 5.3.

The F-measure was considered as prediction quality measure and its value varied from 0.04 to 1.00 depending on the method, dataset, length of the evolution chain and classifier used. The best results for the *SGCI* method were achieved for Salon24 dataset, evolution chains of length 6 and longer, using either RandomForest or AdaBoost (with J48 decision tree) classifiers. The GED method revealed its best predictive ability also on Salon24 dataset, evolution chains of length 7 and the same classifiers. Overall, RandomForest and AdaBoost classifiers were better than the other classifiers, however, the differences were not as significant as at extension of evolution chain length. Therefore, for both methods, the best results were for the Salon24 dataset, where groups last over many time frames and had longer history, RandomForest and AdaBoost classifiers as well as long evolution chains. For the short lengths of evolution chains all classifiers delivered rather poor results.

Usually, the both methods reached their almost maximum quality for certain evolution chain length, depending on the dataset and predicting event type. It was at the level of 3 to 7 last time frames for the GED-based method and 3 to 5 for SGCI. It means that extension of evolution chains beyond this border does not significantly improve prediction quality.

Applying Backward Feature Elimination mechanism allowed to observe that: (1) the quality of the results was still at the same level even if only few (out of hundreds) features were selected; and (2) features were mainly selected from the last three time frames which means that most recent time frames has the biggest impact on its nearest future. After carrying out experiments with feature selection also on DBLP and Salon24 datasets, we can conclude that results varies between datasets, but some measures are more often used than others. In case of GED the most frequent are size and cohesion, but for SGCI–max closeness, min betweenness and size.

Hence, the most important conclusions drawn from the experimental studies are: (1) the longer group history the greater prediction quality; (2) the most recent history of the group most influences on its next change and (3) extension the feature set with the information from periods older than a given threshold does not significantly improve prediction.

Despite a lot of work performed in these studies, there is still a need to continue the research in this area; for example in order to examine other measures to describe group profile, particularly reflecting group dynamics, (e.g., change in size of the group between the following time frames, not only absolute size values) or to validate the results on the greater number of datasets.

## Acknowledgments

The work was partially supported by Fellowship co-financed by European Union within European Social Fund; The European Commission under the 7th Framework Programme, Coordination and Support Action, Grant Agreement Number 316097 [ENGINE]; The National Science Centre the research project 2014-2017 decision no. DEC-2013/09/B/ST6/02317, as well as by the Polish Ministry of Science and Higher Education under AGH–University of Science and Technology Grant 11.11.230.124 (statutory project).

## Author Contributions

All authors conceived and designed the experiments, Stanisław Saganowski and Bogdan Gliwa performed the experiments for the GED method and SGCI method, respectively. Data from the Salon24.pl portal were crawled and prepared by Bogdan Gliwa. All authors analyzed the data, wrote the paper and approved the final version of manuscript.

## Conflicts of Interest

The authors declare no conflict of interest.

## References

1.  Wasserman, S.; Faust, K. *Social Network Analysis: Methods and Applications*; Cambridge University Press: Cambridge, UK, 1994.
2.  Liben-Nowell, D.; Kleinberg, J. The link-prediction problem for social networks. *J. Am. Soc. Inf. Sci.* **2007**, *58*, 1019–1031.

3. Lichtenwalter, R.; Lussier, J.T.; Chawla, N.V. New perspectives and methods in link prediction. In Proceedings of the 16th ACM SIGKDD International Conference on Knowledge Discovery and Data Mining, Washington, DC, USA, 25–28 July 2010; ACM: New York, NY, USA; pp. 243–252.

4. Zheleva, E.; Getoor, L.; Golbeck, J.; Kuter, U. *Using Friendship Ties and Family Circles for Link Prediction, SNAKDD'08*; Springer: Berlin/Heidelberg, Germany, 2008; pp. 97–113.

5. Chiang, K.Y.; Natarajan, N.; Tewari, A.; Dhillon, I.S. Exploiting longer cycles for link prediction in signed networks. In Proceedings of the CIKM 2011, Glasgow, UK, 24–28 October 2011; ACM: New York, NY, USA; pp. 1157–1162.

6. Kunegis, J.; Lommatzsch, A.; Bauckhage, C. The slashdot zoo: Mining a social network with negative edges. In Proceedings of the WWW 2009, Madrid, Spain, 20–24 April 2009; ACM: New York, NY, USA; pp. 741–750.

7. Leskovec, J.; Huttenlocher, D.P.; Kleinberg, J.M. Predicting positive and negative links in online social networks. In Proceedings of the WWW 2010, Raleigh, NC, USA, 26–30 April 2010; ACM: New York, NY, USA; pp. 641–650.

8. Symeonidis, P.; Tiakas, E.; Manolopoulos, Y. Transitive node similarity for link prediction in social networks with positive and negative links. In Proceedings of the RecSys 2010, Barcelona, Spain, 26–30 September 2010; ACM: New York, NY, USA; pp. 183–190.

9. Davis, D.; Lichtenwalter, R.; Chawla, N.V. Supervised methods for multi-relational link prediction. In *Social Network Analysis and Mining*; Springer: Vienna, Austria, 2012; doi:10.1007/s13278-012-0068-6.

10. Richter, Y.; Yom-Tov, E.; Slonim, N. Predicting Customer Churn in Mobile Networks through Analysis of Social Groups. In Proceedings of the SDM 2010, Columbus, OH, USA, 29 April–1 May 2010; pp. 732–741.

11. Wai-Ho, A.; Chan, K.C.C.; Xin, Y. A novel evolutionary data mining algorithm with applications to churn prediction. *IEEE Trans. Evol. Comput.* **2003**, *7*, 532–545.

12. Kairam, S.R.; Wang, D.J.; Leskovec, J. The life and death of online groups: Predicting group growth and longevity. In WSDM'12 Proceedings of the fifth ACM International Conference on Web Search and Data Mining, Seattle, WA, USA, 8–12 February 2012; pp. 673–682.

13. Patil, A.; Liu, J.; Gao, J. Predicting group stability in online social networks. In WWW '13 Proceedings of the 22nd International Conference on World Wide Web, Rio de Janeiro, Brazil, 13–17 May 2013; pp. 1021–1030.

14. Goldberg, M.; Magdon-Ismail, M.; Nambirajan, S.; Thompson, J. Tracking and Predicting Evolution of Social Communities. In Proceedings of Privacy, Security, Risk and Trust (PASSAT) and 2011 IEEE Third International Conference on Social Computing (SocialCom), Boston, MA, USA, 9–11 October 2011; pp. 780–783.

15. Matjaž, P. The Matthew effect in empirical data. *J. R. Soc. Interface* **2014**, *11*, doi:10.1098/rsif.2014.0378.

16. Bródka, P.; Kazienko, P.; Kołoszczyk, B. Predicting Group Evolution in the Social Network. In *Social Informatics*; Aberer, K., Flache, A., Jager, W., Liu, L., Tang, J., Guéret, C., Eds.; Springer: Berlin/Heidelberg, Germany, 2012; pp. 54–67.

17. Gliwa, B.; Bródka, P.; Zygmunt, A.; Saganowski, S.; Kazienko, P.; Koźlak, J. Different Approaches to Community Evolution Prediction in Blogosphere. In Proceedings of 2013 IEEE/ACM International Conference on Advances in Social Networks Analysis and Mining (ASONAM), Niagara Falls, ON, USA, 25–28 August 2013; pp. 1291–1298.

18. Takaffoli, M.; Rabbany, R.; Zaiane, O.R. Community evolution prediction in dynamic social networks. In Proceedings of 2013 12th International Conference on Machine Learning and Applications (ICMLA), Miami, FL, USA, 4–7 December 2013; pp. 191–196.

19. Derényi, I.; Palla, G.; Vicsek, T. Clique Percolation in Random Networks. *Phys. Rev. Lett.* **2005**, *94*, 160–202.

20. Palla, G.; Derényi, I.; Farkas, I.; Vicsek, T. Uncovering the Overlapping Community Structure of Complex Networks in Nature and Society. *Nature* **2005**, *435*, 814–818.

21. Indyk, W.; Kajdanowicz, T.; Kazienko, P. Relational Large Scale Multi-label Classification Method for Video Categorization. *Multimedia Tools Appl.* **2013**, *65*, 63–74.

22. Kajdanowicz, T.; Kazienko, P. Multi-label Classification Using Error Correcting Output Codes. *Int. J. Appl. Math. Comput. Sci.* **2012**, *22*, 829–840.

23. Bródka, P.; Saganowski, S.; Kazienko, P. GED: The Method for Group Evolution Discovery in Social Networks. *Soc. Netw. Anal. Min.* **2013**, *3*, 1–14.

24. Gliwa, B.; Saganowski, S.; Zygmunt, A.; Bródka, P.; Kazienko, P.; Koźlak, J. Identification of Group Changes in Blogosphere. In Proceedings of 2012 IEEE/ACM International Conference on Advances in Social Networks Analysis and Mining (ASONAM), Istanbul, Turkey, 26–29 August 2012; pp. 1201–1206.

25. Zygmunt, A.; Bródka, P.; Kazienko, P.; Koźlak, J. Key person analysis in social communities within the blogosphere. *J. UCS* **2012**, *18*, 577–597.

26. Gliwa, B.; Zygmunt, A.; Byrski, A. Graphical analysis of social group dynamics. In Proceedings of Fourth International Conference on Computational Aspects of Social Networks, CASoN 2012, Sao Carlos, Brazil, 21–23 November 2012; pp. 41–46.

27. Freeman, L.C. Centrality in Social Networks. Conceptual Clarification. *Soc. Netw.* **1978–1979**, *1*, 215–239.

28. Newman, M. *Networks: An Introduction*; Oxford University Press: Oxford, UK, 2010.

29. Bonacich, P.B. Factoring and weighing approaches to status scores and clique identification. *J. Math. Sociol.* **1972**, *2*, 113–120.

30. Ley, M. The DBLP computer science bibliography: Evolution, research issues, perspectives. In *String Processing and Information Retrieval*; Laender, A.F., Oliveira, A., Eds.; Springer: Berlin/Heidelberg, Germany, 2002; pp. 1–10.

31. Viswanath, B.; Mislove, A.; Cha, M.; Gummadi, K.P. On the evolution of user interaction in Facebook. In Proceedings of WOSN '09 Proceedings of the 2nd ACM workshop on Online social networks, Barcelona, Spain, 16–21 August 2009; pp. 37–42.

32. Bródka, P.; Musiał, K.; Kazienko, P. A Performance of Centrality Calculation in Social Networks, CASoN 2009. In Proceedings of CASON '09. International Conference on Computational Aspects of Social Networks, Paris, French, 24–27 June 2009; pp. 24–31.

33. McLachlan, G.J.; Do, K.A.; Ambroise, C. *Analyzing Microarray Gene Expression Data*; John Wiley & Sons: Hoboken, NJ, USA, 2004; ISBN-10: 0471226165.

34. Quinlan, R. *C4.5: Programs for Machine Learning*; Morgan Kaufmann Publishers: San Mateo, CA, USA, 1993.

35. Breiman, L. Random Forests. *Mach. Learn*. **2001**, *45*, 5–32.

36. Freund, Y.; Schapire, R.E. A Decision-Theoretic Generalization of On-Line Learning and an Application to Boosting. *J. Comput. Syst. Sci.* **1997**, *55*, 119–139.

37. Breiman, L. Bagging predictors. *Mach. Learn.* **1996**, *24*, 123–140.